\documentclass[pra,twocolumn,aps,superscriptaddress]{revtex4}
\usepackage{graphics}
\usepackage{color}
\usepackage{graphicx}
\usepackage{bm}
\usepackage[tight]{subfigure}

\newcommand{\uq}{ARC Centre of Excellence for Quantum-Atom Optics, 
Department of Physics, University of Queensland, Brisbane, 
QLD 4072, Australia.}
\newcommand{\otago}{The Jack Dodd and Dan Walls Centre for Photonics and Ultra Cold Atoms, Univesity of Otago, New Zealand} 
\newcommand{\vuw}{School of Chemical and Physical Sciences, Victoria University of Wellington, New Zealand}
\usepackage{graphicx,psfrag}% eps files
\usepackage{bm}% bold math

\newcommand{\etal}{{\em et al.}}

\newcommand{\EQ}[1]{\begin{eqnarray}#1\end{eqnarray}}

\newcommand{\LGP}{\mathcal{L}_{\rm GP}}

\newcommand{\wL}{\omega_{\scriptscriptstyle L}}

\begin{document}
\title{The Quantum de Laval Nozzle: stability and quantum dynamics of sonic horizons in a toroidally trapped Bose gas containing a superflow}
\author{P.~Jain}
\email{piyushnz@gmail.com}
\affiliation{\vuw}
\affiliation{\otago}
\author{A.~S.~Bradley}
\email{abradley@physics.uq.edu.au}
\affiliation{\uq}
\author{C.~W.~Gardiner}
\email{gardiner@physics.otago.ac.nz}
\affiliation{\otago}
%\date{\today}
\begin{abstract}
We study an experimentally realizable system containing stable black hole-white hole acoustic 
horizons in toroidally trapped Bose-Einstein condensates - the quantum de Laval nozzle. 
We numerically obtain stationary flow configurations and assess their stability using
Bogoliubov theory, finding both 
in hydrodynamic and non-hydrodynamic regimes there exist dynamically unstable regions 
associated with the creation of positive and negative energy quasiparticle pairs in analogy 
with the gravitational Hawking effect. The dynamical instability takes the form of a two
mode squeezing interaction between resonant pairs of Bogoliubov modes. 
We study the evolution of dynamically unstable flows using the 
truncated Wigner method, which confirms the two mode squeezed state picture of the 
analogue Hawking effect for low winding number. 
\end{abstract}
\maketitle

%%%%%%%%%%%%%%%%%%%%%%%%%%%%%%%%%%%%%%%%%%%%%%%

\section{Introduction}
The idea of analogue gravity~\cite{Barcelo2006}, which includes the possibility of observing the analogue of the Hawking effect in fluid systems exhibiting sonic horizons, was first put forward by Unruh  \cite{Unruh1981}. In that paper, using an analysis similar to Hawking's original analysis for cosmological black holes \cite{Hawking1974,Hawking1975}, Unruh showed an acoustic black hole should emit sound waves with a Planckian spectrum at the Hawking temperature 
\begin{eqnarray}\label{eq:hawkingtempunruh}
k_B T_H = \frac{\hbar g_H}{2\pi c_H}
\end{eqnarray}
where $g_H$ is the surface gravity at the black hole horizon and $c_H$ is the speed of sound at the horizon.
The derivation required the quantization of a scalar field propagating in a classical fluid, analogous to a classical gravitational field.
In Bose-Einstein condensates (BECs) long wavelength excitations propagate hydrodynamically, giving a direct analogue model in a system where the Hawking temperature is relatively large~\cite{Barcelo2003b}, and potentially measurable using recently proposed schemes based on Raman spectroscopy~\cite{Schutzhold2006}, or parametric resonance~\cite{Modugno2006}. 

%To date, several groups have undertaken theoretical investigations of sonic horizons and the Hawking effect in BECs, most notably:
%{\bf Garay and co-workers \cite{Garay2000, Garay2001}:} 
%{\bf Barcelo and co-workers \cite{Visser2002,Barcelo2003b}:}
%{\bf Leonhardt and co-workers \cite{Leonhardt2003, Leonhardt2003b}:} 
%{\bf Giovanazzi and co-workers \cite{Giovanazzi2004}:} 

\begin{figure}[!htb]\begin{centering}
\includegraphics[width=.8\columnwidth]{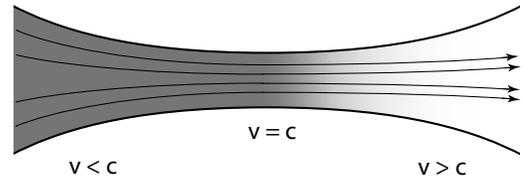}\par\vspace*{-10pt}
\caption{\label{fig:schematic}{\bf Hydrodynamic de Laval nozzle.} When the flow speed (v) of a normal fluid through a constriction achieves the sound velocity (c) at the waist, the outgoing flow becomes supersonic. To maintain continuity, fluid density (grayscale) and stream line spacing (solid lines) decrease from left to right as the fluid accelerates.}
\end{centering}
\vspace*{-10pt}
\end{figure}

%The above investigations notwithstanding, it is still not entirely clear whether the prediction of the Hawking effect is valid for acoustic black hole geometries in quantum systems such as BECs. 
Even in the ultra-cold regime where BECs occur, an extremely low temperature and low losses are both desirable features of any experimental setup to test analogue Hawking radiation in BECs, since it is inherently a delicate and weak signal.
One way to realize a sonic horizon in a trapped BEC without any outcoupling is by setting up a persistent supercurrent in a toroidal trap with ``bumps" in the trapping potential. The bumps act like constrictions, creating a de Laval nozzle~\cite{LandL} type configuration (Fig.~\ref{fig:schematic}).
The experimental realization of a toroidal magnetic trap for ultracold atoms, first demonstrated by using two current-carrying loops \cite{Sauer2001}, has since been developed using magnetic waveguides \cite{Gupta2005}, a microchip trap \cite{Crookston2005} and a four loop configuration \cite{Arnold2006}. The requisite bumps could be introduced optically by a detuned laser shining through an appropriately patterned mask~\cite{Tung2006a}.  
%There have also been proposals for generating persistent currents in toroidal traps~%%\cite{Brand2001,Swingle2005}. 
%---------------------------------------------------------------------------
%----------------------------------------------------------------------------

While there exist previous theoretical studies of de Laval nozzle geometries in the context of acoustic black holes, both for classical fluids  \cite{Sakagami2002,Furuhashi2006} and for BECs \cite{Visser2002,Barcelo2003b}, the analyses in these studies are either classical or semiclassical in nature. 
Previous investigations of acoustic black hole geometries \cite{Barcelo2003b,Visser2002,Leonhardt2003} have focused on regimes where both the hydrodynamical and geometric acoustics descriptions for a BEC apply, so that the semiclassical WKB method can be applied to calculate the Hawking temperature in close analogy with the gravitational derivation~\cite{Hawking1974,Hawking1975,Unruh1981,Visser1993}. 
In particular, flows are treated as hydrodynamic, and the effects of the trap are neglected through some form of local density or WKB approximation. This is understandable given the conditions under which Hawking first discovered the effect, but our primary interest are quantum effects which have also been studied for other BEC acoustic horizon scenarios~\cite{Garay2001,Visser2002,Barcelo2003b,Leonhardt2003, Leonhardt2003b,Giovanazzi2004}. 

In this work, we consider a quite different regime where hydrodynamics is valid only for the low energy modes, whereas geometric acoustics is only valid in the limit of high
winding number. The system of interest is in the region of parameter space where hydrodynamics and geometric acoustics approximations are not usually applicable, but where
progress can be made with more detailed numerical analysis. 

We introduce and analyze a system which exhibits an acoustic black hole and white hole horizon in a trapped BEC, formed by two de Laval nozzles in a toroidal geometry.  
Under the conditions of steady flow this configuration represents the simplest stable de Laval geometry for a Hamiltonian BEC system, and has some appealing properties 
for studying the analogue Hawking effect. 
In particular, it has a discrete excitation spectrum and periodic boundary conditions. Our primary aims are to find stationary solutions of the Gross-Pitaevskii equation for this
system, and to investigate their stability and quantum dynamics. 

\section{Quantum de Laval Nozzle}\label{sect:QdLN}
A weakly interacting Bose gas trapped in a one dimensional potential $V(x)$ at zero temperature is well described by the Gross-Pitaevskii equation~\cite{Dalfovo1999,Garay2001}
\begin{equation}\label{gpelaval}
i\hbar\frac{\partial\psi(x,t)}{\partial t}=\left(-\frac{\hbar^2\partial_x^2}{2m}+V(x)+U_{1{\rm D}}|\psi(x,t)|^2\right)\psi(x,t)
\end{equation}
where the effective one dimensional interaction strength is $U_{1{\rm D}} = U_0/(4\pi r_\perp^2)$ and $U_0=4\pi\hbar^2 a_s/m$ is the usual $s$-wave interaction parameter. The reduction to one dimension assumes the transverse wavefunction is in the harmonic oscillator ground state of toroidal trap: $\phi(r)=\left(1/\pi r_\perp^2\right)^{1/2}e^{-r^2/2r_\perp^2}$, for which we require $\mu\ll \hbar \omega_\perp$, where $\omega_\perp$ is the transverse trapping frequency. The dynamics become effectively one dimensional in this regime since the transverse motion is frozen out by the large energy required to excite transverse modes. The $s$-wave scattering description remains valid provided the scattering length $a$ is much smaller than the transverse dimension $r_{\bot}$ so that the scattering remains effectively three dimensional~\cite{Morgan2002a}. The wavefunction can be written as a macroscopic order parameter
\begin{equation}
\psi(x, t) = \sqrt{n(x, t)} e^{i \Theta(x, t)}
\end{equation}
with current density
\begin{equation}
J \equiv \frac{\hbar}{2 m i} (\psi^* \nabla \psi - \psi \nabla \psi^*)
\end{equation}
and velocity $v = \hbar\,\partial_x \Theta/m$.
In density-phase variables the system is governed by the equations of motion
\begin{equation}\label{gpcontinuity}
\frac{\partial n}{\partial t} + \partial_x (n v) = 0
\end{equation}
and 
\begin{equation}\label{gpeuler}
- \hbar \frac{\partial \Theta}{\partial t} = -\frac{\hbar^2}{2 m \sqrt{n}} \partial_x^2 \sqrt{n}
+ V(x) + U_{1{\rm D}}  n + \frac{1}{2} m v^2.
\end{equation}
When the interaction term dominates, the density varies slowly, and the Laplacian term $\partial_x^2$ (quantum pressure) can be dropped; then the last equation can be takes the form of Euler's equation
\begin{equation}\label{gpeulerhydro}
m \frac{\partial v}{\partial t} = - \partial_x \left( V(x) + U_{1{\rm D}}  n + \frac{1}{2} m v^2 \right)
\end{equation}
and we recover the classical isentropic flow equations.
Combining (\ref{gpcontinuity}) and (\ref{gpeulerhydro}), one can derive the \emph{nozzle equation} \cite{LandL, FluidMechYih}.
\begin{eqnarray}\label{nozzle}
\frac{dv}{v} =  \left( \frac{1}{1 - (v/c)^2} \right)\frac{d V(x)}{m c^2}
\end{eqnarray}
relating variations of the potential and the flow velocity.
The physical consequences of this form of the nozzle equation are as follows (refer to Fig. \ref{fig:schematic}):
Sonic flow ($v = c$) is only permitted where $d V = 0$, that is
at the waist of the nozzle.
For subsonic flow ($v < c$), when $dV$ is negative/positive the velocity is decreasing/increasing. Conversely for supersonic flow, when $dV$ is negative/positive the velocity is increasing/decreasing. Therefore, if the flow is subsonic on approach to the nozzle waist, becoming sonic at the waist, it becomes supersonic on exiting the waist, and conversely for an approaching supersonic flow.

To achieve transonic steady flow in a toroidal geometry, two de Laval nozzles
are required in tandem, the flow becoming supersonic at the waist of the first, and then subsonic
at the waist of the second. This configuration corresponds to the formation of both a black and white hole horizon. 
To implement such a geometry we consider an external potential of the form
\begin{equation}\label{potential}
V(x) = - V_0 \sin^2 \left(\frac{2 \pi x}{L} \right)
\end{equation}
which has periodicity 2 over the interval $-L/2 \leq x \leq L/2$, which is periodic for the toroidal geometry.
For a BEC confined by such a potential, the stationary states are found by solving the time-independent Gross-Pitaevskii equation subject to phase quantization. Since, as we will see below, the solutions exhibit strong modifications to hydrodynamic behavior, we refer to this configuration as the \emph{quantum de Laval nozzle} (QdLN). 

We wish to study currently realistic or potentially achievable parameters. Toroidal ultra-cold atom waveguides have been developed by several 
groups~\cite{Gupta2005,Arnold2006}. We take as nominal values those of the recent experiments of the Stamper-Kurn group~\cite{Gupta2005}. 
The experiments typically consist of $N_0\sim3\times 10^5$ $^{87}$Rb atoms held in a toroidal trap with transverse frequency 
$\omega_\perp\sim 2\pi\;80\;{\rm Hz}$, and radius $R\sim 1\;{\rm mm}$ which gives an azimuthal length $L\sim 6.2\;{\rm mm}$ and a 
transverse harmonic oscillator dimension $r_\perp = \sqrt{\hbar\omega_\perp/m}\sim 1.2\;{\rm \mu m}$. Typical winding numbers are estimated 
by assuming that the circulation velocity is constant and assuming that the gas fills the entire perimeter of the toroid (which is not the case 
in the experiment). This leads to the estimate $w_0=mL^2/2\pi\hbar T$, which for the circulation period of Ref.~\cite{Gupta2005} ($\sim 200 {\rm ms}$) 
gives $w_0\sim 3\times 10^4$ which is quite large. Constructing stationary solutions for the quantum de Laval nozzle for these exact parameters presents 
a major computational challenge as it is necessary to resolve phase variations of the wavefuction on a very small scale, however we note that the 
condensates in this experiment are launched into the toroid and do not form periodic stationary solutions. In theoretical work existing in the 
literature more modest winding numbers have been considered: $1\leq w_0 \lesssim 10$~\cite{Garay2001}. In this work we will examine the stability 
of a similarly modest range of winding numbers in detail, and also for comparison we include $w_0=50$. 

\section{Stationary states}\label{sect:lavalstationary}
In this section we find stationary solutions for the QdLN using the Gross-Pitaevskii equation and compare the results with hydrodynamic and perturbative approaches. 

It is convenient to normalize the condensate wavefunction to the single particle form 
$\int{|\psi(x)|^2\,dx} = 1$ in what follows, so that $g\equiv N_0U_{1{\rm D}}$ hereafter describes the total effective nonlinearity for $N_0$ condensate atoms.
For steady state flow we take the one dimensional stationary solution of the form
\begin{equation}\label{wavefunction1D}
\psi(x, t) = \sqrt{n(x)} \, e^{i \vartheta(x)} e^{- i \mu t / \hbar}.
\end{equation}
If we take the stationary solution of (\ref{gpcontinuity}) and (\ref{gpeuler})
using (\ref{wavefunction1D}), and take the fixed current condition, we can write $
n \equiv s^2$, $v=J/s^2$, 
\begin{eqnarray}
\mu s&=& -{\hbar^2\over 2m}{d^2s\over dx^2}+ V(x)s + g s^3
+ {m J^2\over 2s^3}. \label{eqnmotion}
\end{eqnarray}
Solutions to this nonlinear equation then allow us to reconstruct the wavefunction by specifying:
\begin{eqnarray}
\vartheta(x) &=& {m\over\hbar}\int^x{J\,dy\over s(y)^2},
\\
\psi(x) &=& s(x)\exp\big(i\vartheta(x)\big).
\end{eqnarray}
Solutions must also satisfy the phase quantization condition
\begin{eqnarray}\label{phaseconstr2}
\frac{m}{\hbar} \int^{L/2}_{-L/2} {v(x) dx} =  2 \pi w_0
\end{eqnarray}
for an integer winding number given by $w_0$.

\subsection{Hydrodynamic solutions}\label{sect:hydrodynamic}
When the interactions dominate the density varies slowly so we can invoke the hydrodynamic approximation and drop the Laplacian
term. In this case (\ref{eqnmotion}) can be written as a cubic, either in terms of the density:
\begin{eqnarray}\label{hdcubicdensity}
n^3 + \left( \frac{V(x) - \mu}{g} \right) n^2 + \frac{m J^2}{2 g} = 0
\end{eqnarray}
or in terms of the velocity:
\begin{eqnarray}\label{hdcubicvelocity}
v^3 + 2 \left( \frac{V(x) - \mu}{m} \right) v + \frac{2 J g}{m} = 0.
\end{eqnarray}

%----------------------------------------------------------------------------------
For the case where the flow is zero, there is one non-trivial solution to (\ref{hdcubicdensity})
\begin{equation}
n = \frac{\mu - V(x)}{g}.
\end{equation}
The chemical potential is $\mu = g/L + V_0/2$, which yields the density
\begin{equation}
n = \frac{1}{L} + \frac{V_0}{g} \left( \sin^2\left(\frac{2 \pi x}{L}\right)+\frac{1}{2} \right).
\end{equation}

%----------------------------------------------------------------------------------
On the other hand, for the case where there is non-zero flow ($J> 0$), we find solutions using (\ref{hdcubicvelocity}) since the equations have a simpler form in this case. The solutions are conveniently separated by the discriminant of the cubic~\cite{Abram}
\begin{equation}\label{lavaldiscriminant}
d(x) \equiv \frac{8}{27}\left(\frac{V(x)-\mu}{m}\right)^3+\left(\frac{Jg}{m}\right)^2.
\end{equation}
Transonic configurations exist when there are two real positive solutions, which occurs when $d(x)\leq 0$ for all $x$. Since $(Jg/m)^2 > 0$ the negative semi-definite character of $d(x)$ imposes the constraint $V_0 \leq \mu$.
We can express these solutions analytically as:
\begin{eqnarray}%\label{hydrosolutions}
v_-(x) &=& \sqrt{\frac{8(\mu-V(x))}{3m}}\cos{\left(\frac{\theta(x)+4\pi}{3}\right)}, \label{hydrosolutions2} \\
v_+(x) &=& \sqrt{\frac{8(\mu-V(x))}{3m}}\cos{\left(\frac{\theta(x)}{3}\right)}, \label{hydrosolutions3}
\end{eqnarray}
with 
\begin{equation}\label{hydrosolutions4}
\theta(x)=\cos^{-1}{\left(-\frac{J g}{m}\left[\frac{3 m}{2(\mu - V(x))}\right]^{3/2}\right)}.
\end{equation}
Note $v_-(x)$ is the subsonic branch, whereas $v_+(x)$ is the supersonic branch.

A continuous, single valued transonic solution can be constructed when the two positive solutions coincide at the horizon,  which we take to be $x=x_H=0$.  This occurs when $d(x_H)=0$. 
At the horizon $V(x_H)=0$ (ie. the maximum of the potential acts as the \emph{waist} of the de Laval nozzle), so rearranging (\ref{lavaldiscriminant}) we find the condition for transonic flow is given by the critical chemical potential
\begin{eqnarray}\label{mucrit}
\mu_{{\rm crit}} = \frac{3}{2} (Jg)^{2/3} m^{1/3}.
\end{eqnarray}
Note for $\mu < \mu_{\rm crit}$ we have $d(x) > 0$ and the flow is unstable.

We can find a transonic solution by taking the chemical potential $\mu = \mu_{{\rm crit}}(x_{H})$ so that there is a crossover from the subsonic to supersonic branches at the horizon: $v_-(x_H)=v_+(x_H)$. The transonic solution is then constructed by joining the subsonic ($v_-$) and supersonic ($v_+$) solution branches. Without loss of generality, we take the subsonic branch to span the interval $x \in [-L/2,0]$, with the supersonic branch in the interval $x \in [0,L/2]$.  Consistent solutions with integer winding number $w_0$ are found by iterating the hybrid solutions and the constraints to find the appropriate conserved current $J=n(x)v(x)$.
The resulting stationary solution is fully determined by the parameters $V_0$, $g$ and $w_0$. 

\subsection{Perturbation theory}
Although we have the analytical solutions of the hydrodynamic theory it is useful to adopt a perturbative approach 
which has the advantage of giving simple and reasonably accurate solutions at first order in powers of $\epsilon^{1/2}\equiv (V_0/\mu_0)^{1/2}$, 
where $\mu_0$ is the zeroth order chemical potential. We have chosen our potential so that we can choose the unperturbed 
problem as the homogeneous solution for $V_0=0$, with critical flow so that $v=c$ everywhere. 

At any order the solutions must satisfy the set of equations
\EQ{
v^3-\frac{2}{m}\left(\mu+V_0\sin^2{\left(\frac{2\pi x}{L}\right)}\right)v+\frac{2Jg}{m}&=&0,\\
\frac{8}{27}\left(\frac{\mu}{m}\right)^3-\left(\frac{Jg}{m}\right)^2&=&0,\\
J-nv&=&0,\\
\frac{m}{2\pi\hbar}\int_{-L/2}^{L/2}v\;dx-w_0&=&0.
}
At zeroth order, the potential free equations satisfied by the unperturbed variables $(\mu_0,J_0,v_0)$ are
\EQ{
v_0^3-\frac{2\mu_0v_0}{m}+\frac{2J_0g}{m}&=&0,\\
\frac{8}{27}\left(\frac{\mu_0}{m}\right)^3-\left(\frac{J_0g}{m}\right)^2&=&0,\\
J_0-n_0v_0&=&0,\\
\frac{m}{2\pi\hbar}\int_{-L/2}^{L/2}v_0 dx-w_0&=&0.
} 
In solving the cubic we find solutions $v_0\in(-2(J_0g/m)^{1/3},(J_0g/m)^{1/3},(J_0g/m)^{1/3})$. Only the positive flow solutions are physical and their coalescence at zeroth order is helpful at higher order where the solutions break the parity symmetry of the potential. 

At zeroth order the solutions can expressed in terms of the winding number as
$v_0=2\pi\hbar w_0/mL$, $J_0=mv_0^3/g$, $\mu_0=3mv_0^2/2$, and 
$n_0=mv_0^2/g$. Since we are going to require $V_0\ll\mu_0$ this imposes the condition
\EQ{
\left(\frac{mV_0L^2}{6\pi^2\hbar^2}\right)^{1/2}\ll w_0
}
for the validity of the perturbation series. We introduce the rescaling $(v,J,\mu,x)=(\bar{v}v_0,\bar{J}J_0,\bar{\mu}\mu_0,\bar{x}L)$, to obtain the equations 
\EQ{
\label{vbar}\bar{v}^3-3\bar{v}(\bar{\mu}+\epsilon\sin^2(2\pi \bar{x}))+2\bar{J}&=&0,\\
\label{crit}\bar{\mu}^3-\bar{J}^2&=&0,\\
\label{Jbar}\bar{J}-\bar{n}\bar{v}&=&0,\\
\label{w0}\int_{-1/2}^{1/2} \bar{v}\;d\bar{x}-1&=&0.
}
The repeated solution at zeroth order means we have to use a perturbation series in powers of $\epsilon^{1/2}$~\cite{Bush}, so we assume an expansion of the form $
\bar{v}=1+\epsilon^{1/2}v_1+\epsilon v_2\dots$ and similarly for the other variables. We can obtain consistent solutions to Eqs.~(\ref{vbar}--\ref{w0}) up to $O(\epsilon)$ which give a good qualitative description of the solutions and are quite accurate for a wide range of parameters.

Terms in the expansion of (\ref{vbar}) of order $\epsilon^0, \epsilon^{1/2}$ cancel, and the $\epsilon$ equation is
\EQ{
v_1^2=\mu_1+\sin^2{(2\pi \bar{x})}.
}
Substituting the series into (\ref{crit}) gives $\mu_1=0$ which is not surprising, and in fact $\mu_2=0$.
The subsonic and supersonic solutions are automatically matched at the acoustic horizon by choosing $v_1=\sin{(2\pi \bar{x})}$, whereby the supersonic region is $0<\bar{x}<1/2$, coinciding with our previous hydrodynamic treatment. 

Returning to dimensioned variables, the first order solutions for the quantum de Laval nozzle are
\EQ{
v&=&v_0\left(1+\left(\frac{V_0}{\mu_0}\right)^{1/2}\sin{\left(\frac{2\pi x}{L}\right)}\right)+O\left(\frac{V_0}{\mu_0}\right),\\
n&=&n_0\left(1-\left(\frac{V_0}{\mu_0}\right)^{1/2}\sin{\left(\frac{2\pi x}{L}\right)}\right)+O\left(\frac{V_0}{\mu_0}\right),\;\;
}
with $J=J_0+O((V_0/\mu_0)^{3/2})$ and $\mu=\mu_0+O((V_0/\mu_0)^{3/2})$. These expressions give a good qualitative description of the Gross-Pitaevskii solutions, and are typically very close to the full hydrodynamic solutions (see Fig. \ref{fig:solution_w3}). 
As expected, the differences are more apparent with increasing $\epsilon$ corresponding to a more important quantum pressure term, and with increasing distance from the sonic horizons.

%-----------------------------------------------------------
\subsection{Solutions of the Gross-Pitaevskii equation}\label{sect:eigenstates}
%-----------------------------------------------------------

We now use the transonic solution to the hydrodynamic problem (Sec.~\ref{sect:hydrodynamic}) as a starting point to finding the stationary solutions of the Gross-Pitaevskii equation (GPE). The full numerical solutions exhibit ``ripple'' structures in regions where the quantum pressure term becomes important, in particular, in the region of supersonic flow downstream of the black hole acoustic horizon.

In order to find stationary solutions we use constrained optimization. Imaginary time evolution is not feasible in this case because we are interested in stationary states which are excited into circular motion relative to the ground state. 
We formulate the problem as the the minimization of the Gross-Pitaevski functional for a fixed nonlinearity $g$, potential depth $V_0$ and current $J$, subject to a phase quantization constraint in terms of a fixed winding number $w_0$.
The problem is recast as the set of algebraic equations:
\begin{eqnarray}\label{optim_function}
\left [\frac{\hbar^2}{2m} \frac{d^2}{dx^2} + \mu - V(x) - g s(x)^2 -
\frac{mJ^2}{2s(x)^4} \right] s(x) & = & 0, \nonumber \\
\frac{m}{\hbar}\int_{-L}^{+L} \frac{J}{s(x)^2} \, d x -  2 \pi w & = & 0. \nonumber
\end{eqnarray}
The phase circulation constraint ensures the wavefunction $\psi = s(x) e^{i\vartheta(x)}$ is everywhere single-valued. 

The solution for our vector of unknowns $\mathbf{X} = \{s(x_i), \mu\}$ is found by Levenberg-Marquardt optimization~\cite{Levenberg1944a,Marquardt1963a} in MATLAB using the hydrodynamic solution as the initial condition $\mathbf{X}_0$. The unit of energy for this system which we will use to display our results is $\hbar\omega_L\equiv \hbar^2/(mL^2)$.

% This file is generated by the MATLAB m-file laprint.m. It can be included
% into LaTeX documents using the packages graphicx, color and psfrag.
% It is accompanied by a postscript file. A sample LaTeX file is:
%    \documentclass{article}\usepackage{graphicx,color,psfrag}
%    \begin{document}\input{lavalstate_V0100_w3}\end{document}
% See http://www.mathworks.de/matlabcentral/fileexchange/loadFile.do?objectId=4638
% for recent versions of laprint.m.
%
% created by:           LaPrint version 3.16 (13.9.2004)
% created on:           09-Feb-2007 08:41:32
% eps bounding box:     15 cm x 10.9091 cm
% comment:              
%
\begin{psfrags}%
\psfragscanon%
%
% text strings:
\psfrag{s09}[b][b]{\color[rgb]{0,0,0}\setlength{\tabcolsep}{0pt}\begin{tabular}{c}$|\psi|^2 L$\end{tabular}}%
\psfrag{s14}[][]{\color[rgb]{0,0,0}\setlength{\tabcolsep}{0pt}\begin{tabular}{c} \end{tabular}}%
\psfrag{s15}[][]{\color[rgb]{0,0,0}\setlength{\tabcolsep}{0pt}\begin{tabular}{c} \end{tabular}}%
\psfrag{s16}[l][l]{\color[rgb]{0,0,0}\setlength{\tabcolsep}{0pt}\begin{tabular}{l}(a)\end{tabular}}%
\psfrag{s17}[l][l]{\color[rgb]{0,0,0}hydrodynamic}%
\psfrag{s18}[l][l]{\color[rgb]{0,0,0}GPE}%
\psfrag{s19}[l][l]{\color[rgb]{0,0,0}perturbation}%
\psfrag{s20}[l][l]{\color[rgb]{0,0,0}hydrodynamic}%
\psfrag{s21}[t][t]{\color[rgb]{0,0,0}\setlength{\tabcolsep}{0pt}\begin{tabular}{c}$x/L$\end{tabular}}%
\psfrag{s22}[b][b]{\color[rgb]{0,0,0}\setlength{\tabcolsep}{0pt}\begin{tabular}{c}$v/L\wL$\end{tabular}}%
\psfrag{s26}[][]{\color[rgb]{0,0,0}\setlength{\tabcolsep}{0pt}\begin{tabular}{c} \end{tabular}}%
\psfrag{s27}[][]{\color[rgb]{0,0,0}\setlength{\tabcolsep}{0pt}\begin{tabular}{c} \end{tabular}}%
\psfrag{s28}[l][l]{\color[rgb]{0,0,0}\setlength{\tabcolsep}{0pt}\begin{tabular}{l}BH\end{tabular}}%
\psfrag{s29}[l][l]{\color[rgb]{0,0,0}\setlength{\tabcolsep}{0pt}\begin{tabular}{l}WH\end{tabular}}%
\psfrag{s30}[l][l]{\color[rgb]{0,0,0}\setlength{\tabcolsep}{0pt}\begin{tabular}{l}(b)\end{tabular}}%
\psfrag{s31}[l][l]{\color[rgb]{0,0,0}flow}%
\psfrag{s32}[l][l]{\color[rgb]{0,0,0}sound}%
\psfrag{s33}[l][l]{\color[rgb]{0,0,0}flow}%
%
% xticklabels:
\psfrag{x01}[t][t]{-0.5}%
\psfrag{x02}[t][t]{0}%
\psfrag{x03}[t][t]{0.5}%
%
% yticklabels:
\psfrag{v01}[r][r]{15}%
\psfrag{v02}[r][r]{25}%
\psfrag{v03}[r][r]{0.5}%
\psfrag{v04}[r][r]{1}%
\psfrag{v05}[r][r]{1.5}%
%
% Figure:
%\resizebox{12cm}{!}{\includegraphics{lavalstate_V0100_w3.eps}}%
\begin{figure}[!t]
    \begin{center}
		\includegraphics[width=1\columnwidth]{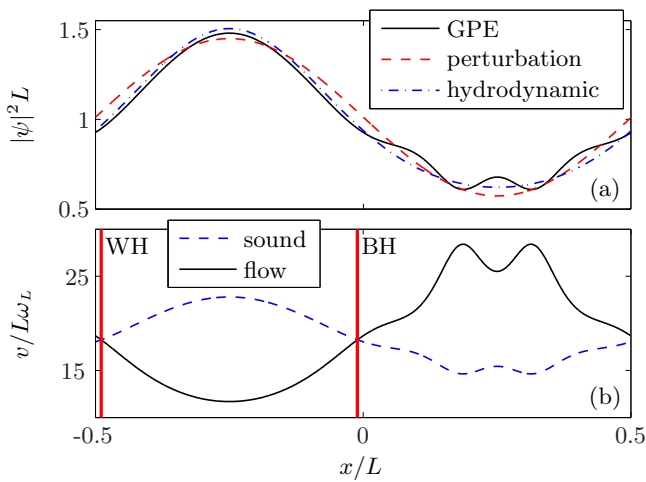}
		\caption{(color online) \textbf{QdLN stationary flow.} Parameters are $w_0  = 3$, $V_0 = 100\hbar\omega_L$ (energies are in units of $\hbar\omega_L\equiv\hbar^2/mL^2$), $\mu = 5.93 \times 10^2\hbar\omega_L$ and $g = 3.51 \times 10^2 L\hbar\omega_L$. (a) Comparison of solutions for condensate density. (b) Velocity and speed of sound for GPE solution.} 
		\label{fig:solution_w3}
		\end{center}
\end{figure}
\end{psfrags}%
%
% End lavalstate_V0100_w3.tex

% This file is generated by the MATLAB m-file laprint.m. It can be included
% into LaTeX documents using the packages graphicx, color and psfrag.
% It is accompanied by a postscript file. A sample LaTeX file is:
%    \documentclass{article}\usepackage{graphicx,color,psfrag}
%    \begin{document}\input{lavalstate_V0100_w10}\end{document}
% See http://www.mathworks.de/matlabcentral/fileexchange/loadFile.do?objectId=4638
% for recent versions of laprint.m.
%
% created by:           LaPrint version 3.16 (13.9.2004)
% created on:           09-Feb-2007 08:48:20
% eps bounding box:     15 cm x 10.9091 cm
% comment:              
%
\begin{psfrags}%
\psfragscanon%
%
% text strings:
\psfrag{s09}[b][b]{\color[rgb]{0,0,0}\setlength{\tabcolsep}{0pt}\begin{tabular}{c}$|\psi|^2 L$\end{tabular}}%
\psfrag{s14}[][]{\color[rgb]{0,0,0}\setlength{\tabcolsep}{0pt}\begin{tabular}{c} \end{tabular}}%
\psfrag{s15}[][]{\color[rgb]{0,0,0}\setlength{\tabcolsep}{0pt}\begin{tabular}{c} \end{tabular}}%
\psfrag{s16}[l][l]{\color[rgb]{0,0,0}\setlength{\tabcolsep}{0pt}\begin{tabular}{l}(a)\end{tabular}}%
\psfrag{s17}[l][l]{\color[rgb]{0,0,0}hydrodynamic}%
\psfrag{s18}[l][l]{\color[rgb]{0,0,0}GPE}%
\psfrag{s19}[l][l]{\color[rgb]{0,0,0}perturbation}%
\psfrag{s20}[l][l]{\color[rgb]{0,0,0}hydrodynamic}%
\psfrag{s21}[t][t]{\color[rgb]{0,0,0}\setlength{\tabcolsep}{0pt}\begin{tabular}{c}$x/L$\end{tabular}}%
\psfrag{s22}[b][b]{\color[rgb]{0,0,0}\setlength{\tabcolsep}{0pt}\begin{tabular}{c}$v/L\wL$\end{tabular}}%
\psfrag{s26}[][]{\color[rgb]{0,0,0}\setlength{\tabcolsep}{0pt}\begin{tabular}{c} \end{tabular}}%
\psfrag{s27}[][]{\color[rgb]{0,0,0}\setlength{\tabcolsep}{0pt}\begin{tabular}{c} \end{tabular}}%
\psfrag{s28}[l][l]{\color[rgb]{0,0,0}\setlength{\tabcolsep}{0pt}\begin{tabular}{l}BH\end{tabular}}%
\psfrag{s29}[l][l]{\color[rgb]{0,0,0}\setlength{\tabcolsep}{0pt}\begin{tabular}{l}WH\end{tabular}}%
\psfrag{s30}[l][l]{\color[rgb]{0,0,0}\setlength{\tabcolsep}{0pt}\begin{tabular}{l}(b)\end{tabular}}%
\psfrag{s31}[l][l]{\color[rgb]{0,0,0}flow}%
\psfrag{s32}[l][l]{\color[rgb]{0,0,0}sound}%
\psfrag{s33}[l][l]{\color[rgb]{0,0,0}flow}%
%
% xticklabels:
\psfrag{x01}[t][t]{-0.5}%
\psfrag{x02}[t][t]{0}%
\psfrag{x03}[t][t]{0.5}%
%
% yticklabels:
\psfrag{v01}[r][r]{60}%
\psfrag{v02}[r][r]{70}%
\psfrag{v03}[r][r]{0.9}%
\psfrag{v04}[r][r]{1}%
\psfrag{v05}[r][r]{1.1}%
%
% Figure:
%\resizebox{12cm}{!}{\includegraphics{lavalstate_V0100_w10.eps}}%
\begin{figure}[!t]
    \begin{center}
		\includegraphics[width=1.0\columnwidth]{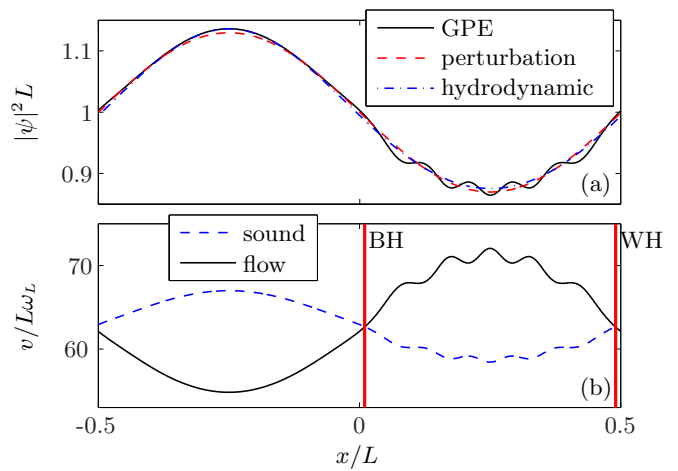}
		\caption{(color online) \textbf{QdLN stationary flow}. Parameters are as in Fig.~\ref{fig:solution_w3}, but with $w_0  = 10$, $\mu = 5.99 \times 10^3\hbar\omega_L$ and $g = 3.95 \times 10^3 L\hbar\omega_L$.} 
		\label{fig:solution_w10}
		\end{center}
\end{figure}
\end{psfrags}%
%
% End lavalstate_V0100_w10.tex

To consider some examples we use a potential with $V_0 = 100\hbar\omega_L$, and show solutions for winding numbers $w_0=3$ (Fig.~\ref{fig:solution_w3}) and $w_0=10$ (Fig.~\ref{fig:solution_w10}). In each case, the full hydrodynamic, first order perturbation theory and GPE solutions are shown, as well as the flow velocity and speed of sound for the GPE stationary solution, and the location of the acoustic black hole (BH) and white hole (WH) horizons. The ergoregion ($v > c$) is given approximately by the right hand region $0 \leq x \leq 0.5$. By construction, in the hydrodynamic case, the black hole horizon occurs at $x_{\rm BH} = 0$, whereas the white hole horizon occurs at $x_{\rm WH} = \pm 0.5$. 

In principle we might expect to be able to vary the winding number $w_0$ and potential depth $V_0$ to find a continuous range of transonic solutions for the QdLN. This is the case, for example, for the toroidal system considered by Garay {\em et al.}~\cite{Garay2001}, where for a given $w_0$ a stability diagram over a continuous range of $V_0$ and $g$ was mapped out. In fact we find that for the QdLN the total nonlinearity $g$ is not a free parameter of the solutions -- it is uniquely determined for a given $(w_0,V_0)$. This is physically reasonable because $w_0$ sets the flow velocity and the nonlinearity determines the speed of sound, and the two must be equal at the sonic horizons. 

We can easily find a very accurate relation between nonlinearity and winding number using the zeroth order perturbation theory: $g=L\hbar\omega_{L}(2\pi w_0)^2$. We should expect deviations from this relationship for low winding number due to the importance of the quantum pressure term, however we find they are very small. Numerically we find for the GPE solutions that $g$ varies according to this quadratic law and depends only very weakly on $V_0$. For the values of $(w_0,V_0)$ used in this paper the behavior is essentially independent of $V_0$ and shows a maximum deviation from the zeroth order perturbation theory result of order $1\%$, occurring at our lowest winding number, $w_0=3$.

Qualitatively, the main departure of the hydrodynamic solutions from first order perturbation theory is a loss of parity: the increasingly important interaction energy eventually lifts the antisymmetry of $v_1 = \sin{(2\pi x/L)}$; such differences are more pronounced for low winding number.

\section{Quasiparticles and stability}\label{sect:lavalbog}
We will now determine the stability of the GPE stationary solutions by finding their Bogoliubov excitation spectra for a wide range of potential depths and winding numbers. For unstable configurations the standard Bogoliubov analysis is insufficient, and we use the theory of Leonhardt {\em et al.}~\cite{Leonhardt2003} to obtain the correct normalizable Bogoliubov modes. These modes show some interesting localization properties with respect to the acoustic horizons, and will be used to construct Bogoliubov vacuum states when we come to dynamical simulations in Sec.~\ref{sect:lavaldynamics}.
\subsection{Normalizable Bogoliubov modes}\label{lavalbdg}
The linear excitations of the condensate are described by the Bogoliubov-de Gennes (BdG) equations~\cite{Morgan1998,STR}. Consider a solution with small oscillations around a stationary state 
\begin{eqnarray}\label{eq:lavalbogexpansion}
\!\!\!\!\psi(x, t) &=& e^{- i \mu t/\hbar} \Big( \phi_0(x)\nonumber\\
&&+\!\sum_i \left[ u_i(x) \beta_i e^{-i \omega_i t} + v_i^*(x) \beta_i^* e^{i \omega_i t} \right] \Big)
\end{eqnarray}
where $\beta_i$ and $\beta_i^*$ are the amplitudes for the oscillations and $\phi_0(x)$ is normalized to unity in what follows. For the quantum field, these quantities are replaced by the bosonic annihilation and creation operators for the excitations, given by $\hat{b}_i$ and $\hat{b}_i^{\dagger}$ respectively, with $[\hat{b}_i,\hat{b}_j^\dag]=\delta_{ij}$. $\phi_0(x)$ is the solution to the time-independent GPE given by (\ref{gpelaval}). 
To obtain the correct physical modes it is necessary to introduce the operator
\begin{eqnarray}
\hat{Q} = 1 - |\phi_0 \rangle \langle \phi_0 | \label{bdgproj1} 
\end{eqnarray}
which projects orthogonal to the condensate. Introducing projectors appropriately and substituting the Bogoliubov expansion into the time-dependent GPE and while only keeping terms linear in $u_i(x)$ and $v_i(x)$ yields the modified BdG equations
\begin{eqnarray}\label{bdgev}
\mathcal{L} \left[ {\begin{array}{*{20}c}
   u_i(x)  \\
   v_i(x)  \\
\end{array}} \right] = \epsilon_i \left[ {\begin{array}{*{20}c}
   u_i(x)  \\
   v_i(x)  \\
\end{array}} \right]
\end{eqnarray}
where the operator $\mathcal{L}$ is given by
\begin{eqnarray}\label{bdgLmatrix}
\mathcal{L}\!\equiv\!\left[\!{\begin{array}{*{20}c}
   \!{\LGP\!-\!\mu\!+\!g \hat{Q} |\phi_0|^2 \hat{Q}} & \!\!\!\!{g \hat{Q} \phi_0^2\hat{Q}^*}  \\
   {-\!g \hat{Q}^* \phi_0^{* 2} \hat{Q}} & \!\!\!\!{-\!\left(\!\LGP\!-\!\mu\!+\!g\hat{Q}|\phi_0|^2\hat{Q}\!\right)^*}\\
\end{array}}\!\right]\nonumber
\end{eqnarray}
and the Gross-Pitaevskii operator is
\begin{eqnarray}
\LGP \equiv - \frac{\hbar^2 \partial_x^2}{2m} + V(x) + g |\phi_0|^2.
\end{eqnarray}
The stationary solution satisfies $(\LGP - \mu) \phi_0 = 0$.
The solutions to this equation are the eigenvalues $\epsilon_i\equiv
\hbar(\omega_i+i\gamma_i)$, and the normal modes of the system. 

The orthogonality and symmetry relations are fixed by the requirement that the many body Hamiltonian for the interacting Bose gas is diagonal (to quadratic order) in quasiparticle operators and that the transformation to quasiparticles preserves the commutation relations. 

The operator $\mathcal{L}$ is not Hermitian so that complex eigenvalues are allowed, corresponding to dynamical instabilities of the system. It is straightforward to show that the modes with complex eigenvalues have zero norm \cite{Fetter1972} and cannot therefore be associated with bosonic operators in the field expansion (\ref{eq:lavalbogexpansion}). However, following Leonardt \etal~\cite{Leonhardt2003}, it is still possible to construct normalizable modes for the unstable modes by the construction:
\begin{eqnarray}\label{eq:bogmodes1}
\left[ {\begin{array}{*{20}c}
   U_i^+(x)  \\
   V_i^+(x)  \\
\end{array}} \right] &=& \frac{1}{\sqrt{2}} \left( \left[ {\begin{array}{*{20}c}
   u_i^+(x)  \\
   v_i^+(x)  \\
\end{array}} \right] + \left[ {\begin{array}{*{20}c}
   v_i^-(x)  \\
   u_i^-(x)  \\
\end{array}} \right]^* \right), \\ \label{eq:bogmodes2}
\left[ {\begin{array}{*{20}c}
   U_i^-(x)  \\
   V_i^-(x)  \\
\end{array}} \right] &=& \frac{1}{\sqrt{2}} \left( \left[ {\begin{array}{*{20}c}
   u_i^-(x)  \\
   v_i^-(x)  \\
\end{array}} \right] -  \left[ {\begin{array}{*{20}c}
   v_i^+(x)  \\
   u_i^+(x)  \\
\end{array}} \right]^* \right)
\end{eqnarray}
where $[u_i^+(x), v_i^+(x)]$ and $[u_i^-(x), v_i^-(x)]$ are the eigenvectors
associated with the unstable positive and  negative energy eigenvalues
respectively. Note that due to the symmetries of (\ref{bdgLmatrix}) we have
$\omega_i^{-} = - \omega_i^{+}$. Hereafter we use $U_i(x)$ and $V_i(x)$ for the
full set of orthonormal modes. The quadratic Hamiltonian for the stable modes takes the standard form for independent harmonic oscillators.
The creation and annihilation operators for the new modes satisfy the commutation relations for bosonic operators, but show up as non-diagonal terms in the Hamiltonian subspace for the dynamically unstable modes as \cite{Leonhardt2003}
%Garay2001, Mine2006}
\begin{eqnarray}\label{eq:bdgH2amplifier}
\hat{H}_{2}&=&\sum_j \hbar\omega_j \Big[ (\hat{b}^{\dagger}_{j+} \hat{b}_{j+}\!-\!\hat{b}^{\dagger}_{j-} \hat{b}_{j-})\nonumber\\
&&-\int d x \, (|V_j^{+}|^2\!-\!|V_j^{-}|^2) \Big] \nonumber \\ 
&&+\!\sum_j i \,\hbar\gamma_j \Big[(\hat{b}_{j+} \hat{b}_{j-}\!-\!\hat{b}^{\dagger}_{j+} \hat{b}^{\dagger}_{j+})\nonumber\\
&&+\int d x \, (U_j^{+} V_j^{-} - U_j^{+*} V_j^{-*})\Big]
\end{eqnarray}
where the sum is taken over only the dynamically unstable modes, and where $\hat{b}_{j\pm}$ is the annihilation operator and $\hat{b}_{j\pm}^{\dagger}$ the creation operator corresponding to the normalizable modes (\ref{eq:bogmodes1}) and (\ref{eq:bogmodes2}). 
For stable modes the hamiltonian reduces to the usual diagonal Bogoliubov form $\hat{H}_2=\sum_j \hbar\omega_j (\hat{b}^{\dagger}_{j} \hat{b}_{j}+1/2-\int dx\;|V_j|^2)$. Dynamically unstable modes are therefore associated with non-degenerate parametric amplification \cite{QO}, which leads to growth in the unstable modes at the expense of the condensate mode. 
For short time dynamics, the complex eigenvalue will generate exponential growth in each unstable mode. It is this effect that has been suggested to provide the closest analogy with the Hawking effect for BECs \cite{Garay2001, Leonhardt2003}. However, this picture neglects higher order interactions that may be present in the full Hamiltonian, and therefore is likely to fail for dynamics on long time scales. We will investigate this further in Sec.~\ref{sect:lavaldynamics}.

\subsection{Stability and mode structure}\label{sect:lavalspectrum}
For a dynamically unstable configuration, we construct normalizable modes using the procedure outlined in Sec.~\ref{lavalbdg}. We additionally sort the eigenvalues in ascending
order by $\omega_i$ and label the modes accordingly.

% This file is generated by the MATLAB m-file laprint.m. It can be included
% into LaTeX documents using the packages graphicx, color and psfrag.
% It is accompanied by a postscript file. A sample LaTeX file is:
%    \documentclass{article}\usepackage{graphicx,color,psfrag}
%    \begin{document}\input{eigenvalues_w3}\end{document}
% See http://www.mathworks.de/matlabcentral/fileexchange/loadFile.do?objectId=4638
% for recent versions of laprint.m.
%
% created by:           LaPrint version 3.16 (13.9.2004)
% created on:           08-Feb-2007 15:34:28
% eps bounding box:     15 cm x 11.25 cm
% comment:              
%
\begin{psfrags}%
\psfragscanon%
%
% text strings:
\psfrag{s03}[b][b]{\color[rgb]{0,0,0}\setlength{\tabcolsep}{0pt}\begin{tabular}{c}$|\omega_i|/\wL$\end{tabular}}%
\psfrag{s04}[t][t]{\color[rgb]{0,0,0}\setlength{\tabcolsep}{0pt}\begin{tabular}{c}\\$|\gamma_i|/\wL$\end{tabular}}%
\psfrag{s05}[l][l]{\color[rgb]{0,0,0}\setlength{\tabcolsep}{0pt}\begin{tabular}{l}(a)\end{tabular}}%
\psfrag{s08}[t][t]{\color[rgb]{0,0,0}\setlength{\tabcolsep}{0pt}\begin{tabular}{c}mode\end{tabular}}%
\psfrag{s09}[b][b]{\color[rgb]{0,0,0}\setlength{\tabcolsep}{0pt}\begin{tabular}{c}$|\omega_i|/\wL$\end{tabular}}%
\psfrag{s10}[t][t]{\color[rgb]{0,0,0}\setlength{\tabcolsep}{0pt}\begin{tabular}{c}\\$|\gamma_i|/\wL$\end{tabular}}%
\psfrag{s11}[l][l]{\color[rgb]{0,0,0}\setlength{\tabcolsep}{0pt}\begin{tabular}{l}(b)\end{tabular}}%
%
% xticklabels:
\psfrag{x01}[t][t]{0}%
\psfrag{x02}[t][t]{0.1}%
\psfrag{x03}[t][t]{0.2}%
\psfrag{x04}[t][t]{0.3}%
\psfrag{x05}[t][t]{0.4}%
\psfrag{x06}[t][t]{0.5}%
\psfrag{x07}[t][t]{0.6}%
\psfrag{x08}[t][t]{0.7}%
\psfrag{x09}[t][t]{0.8}%
\psfrag{x10}[t][t]{0.9}%
\psfrag{x11}[t][t]{1}%
\psfrag{x12}[t][t]{1}%
\psfrag{x13}[t][t]{2}%
\psfrag{x14}[t][t]{3}%
\psfrag{x15}[t][t]{4}%
\psfrag{x16}[t][t]{5}%
\psfrag{x17}[t][t]{6}%
\psfrag{x18}[t][t]{7}%
\psfrag{x19}[t][t]{8}%
\psfrag{x20}[t][t]{9}%
\psfrag{x21}[t][t]{10}%
\psfrag{x22}[t][t]{1}%
\psfrag{x23}[t][t]{2}%
\psfrag{x24}[t][t]{3}%
\psfrag{x25}[t][t]{4}%
\psfrag{x26}[t][t]{5}%
\psfrag{x27}[t][t]{6}%
\psfrag{x28}[t][t]{7}%
\psfrag{x29}[t][t]{8}%
\psfrag{x30}[t][t]{9}%
\psfrag{x31}[t][t]{10}%
%
% yticklabels:
\psfrag{v01}[r][r]{0}%
\psfrag{v02}[r][r]{0.1}%
\psfrag{v03}[r][r]{0.2}%
\psfrag{v04}[r][r]{0.3}%
\psfrag{v05}[r][r]{0.4}%
\psfrag{v06}[r][r]{0.5}%
\psfrag{v07}[r][r]{0.6}%
\psfrag{v08}[r][r]{0.7}%
\psfrag{v09}[r][r]{0.8}%
\psfrag{v10}[r][r]{0.9}%
\psfrag{v11}[r][r]{1}%
\psfrag{v12}[l][l]{$10^{-10}$}%
\psfrag{v13}[l][l]{$10^{-5}$}%
\psfrag{v14}[l][l]{$10^{0}$}%
\psfrag{v15}[r][r]{0}%
\psfrag{v16}[r][r]{200}%
\psfrag{v17}[r][r]{400}%
\psfrag{v18}[r][r]{600}%
\psfrag{v19}[l][l]{$10^{-10}$}%
\psfrag{v20}[l][l]{$10^{-5}$}%
\psfrag{v21}[l][l]{$10^{0}$}%
\psfrag{v22}[r][r]{0}%
\psfrag{v23}[r][r]{200}%
\psfrag{v24}[r][r]{400}%
\psfrag{v25}[r][r]{600}%
%
% Figure:
%\resizebox{12cm}{!}{\includegraphics{eigenvalues_w3.eps}}%
\begin{figure}[!t]
    \begin{centering}
		\includegraphics[width=0.9\columnwidth]{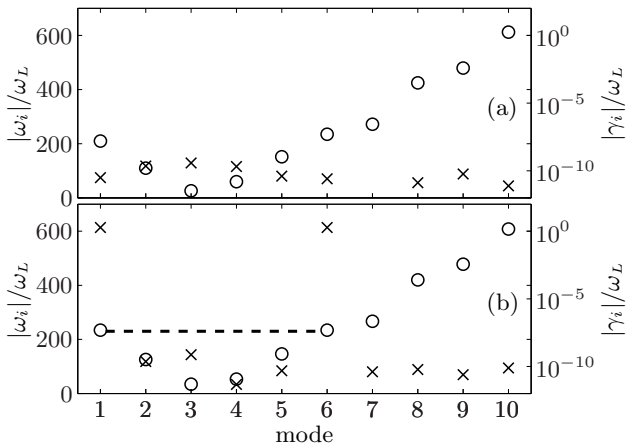}\par
		\end{centering}
		\vspace*{-10pt}
\caption{\label{fig:spec1}\textbf{Eigenspectrum:} We show $\epsilon_i=\hbar(\omega_i+i\gamma_i)$ for $w_0=3$. Circles are for the left axes, and modes are numbered in ordered of increasing $\omega_i$. Mode 4 is the first
positive frequency mode. (a) Stable state for $V_0=140 \hbar\wL$. (b) Dynamically unstable state for $V_0=163.7 \hbar\wL$; the dashed line denotes the mode pair with
$\omega_1=-\omega_6$ signaling the dynamical instability.}
		\vspace*{-10pt}
\end{figure}
\end{psfrags}%
%
% End eigenvalues_w3.tex

%===========================================

Figure \ref{fig:spec1} shows the eigenvalue spectrum for the first few modes with $w_0 = 3$ and two different values of $V_0$: (a) $V_0 = 140 \hbar \wL$; and (b)  $V_0 = 163.7
\hbar \wL$. While both cases have negative eigenvalues, indicating energetic (Landau) instabilities due to non-zero flow, only case (b) exhibits dynamical instabilities also.
Following the theory of the previous section, the onset of a dynamical instability is associated with a pair of modes (labelled by $j$ and $k$ say) with complex eigenvalues that
satisfy $\omega_j = - \omega_k$. In particular, case (b) indicates that modes 1 and 6 are unstable with $\omega_1 = - \omega_6$.

% This file is generated by the MATLAB m-file laprint.m. It can be included
% into LaTeX documents using the packages graphicx, color and psfrag.
% It is accompanied by a postscript file. A sample LaTeX file is:
%    \documentclass{article}\usepackage{graphicx,color,psfrag}
%    \begin{document}\input{laval_stability}\end{document}
% See http://www.mathworks.de/matlabcentral/fileexchange/loadFile.do?objectId=4638
% for recent versions of laprint.m.
%
% created by:           LaPrint version 3.16 (13.9.2004)
% created on:           07-Feb-2007 16:46:32
% eps bounding box:     15 cm x 26.5 cm
% comment:              
%
\begin{psfrags}%
\psfragscanon%
%
% text strings:
\psfrag{s35}[l][l]{\color[rgb]{0,0,0}\setlength{\tabcolsep}{0pt}\begin{tabular}{l}$w_0 = 50$\end{tabular}}%
\psfrag{s36}[l][l]{\color[rgb]{0,0,0}\setlength{\tabcolsep}{0pt}\begin{tabular}{l}$w_0 = 3$\end{tabular}}%
\psfrag{s37}[t][t]{\color[rgb]{0,0,0}\setlength{\tabcolsep}{0pt}\begin{tabular}{c}$V_0/\hbar \wL$\end{tabular}}%
\psfrag{s38}[l][l]{\color[rgb]{0,0,0}\setlength{\tabcolsep}{0pt}\begin{tabular}{l}$w_0 = 4$\end{tabular}}%
\psfrag{s39}[l][l]{\color[rgb]{0,0,0}\setlength{\tabcolsep}{0pt}\begin{tabular}{l}$w_0 = 5$\end{tabular}}%
\psfrag{s40}[l][l]{\color[rgb]{0,0,0}\setlength{\tabcolsep}{0pt}\begin{tabular}{l}$w_0 = 6$\end{tabular}}%
\psfrag{s41}[l][l]{\color[rgb]{0,0,0}\setlength{\tabcolsep}{0pt}\begin{tabular}{l}$w_0 = 7$\end{tabular}}%
\psfrag{s42}[b][b]{\color[rgb]{0,0,0}\setlength{\tabcolsep}{0pt}\begin{tabular}{c}$\textrm{max} |\gamma_i|/\wL$\end{tabular}}%
\psfrag{s43}[l][l]{\color[rgb]{0,0,0}\setlength{\tabcolsep}{0pt}\begin{tabular}{l}$w_0 = 8$\end{tabular}}%
\psfrag{s44}[l][l]{\color[rgb]{0,0,0}\setlength{\tabcolsep}{0pt}\begin{tabular}{l}$w_0 = 9$\end{tabular}}%
\psfrag{s45}[l][l]{\color[rgb]{0,0,0}\setlength{\tabcolsep}{0pt}\begin{tabular}{l}$w_0 = 10$\end{tabular}}%
%
% xticklabels:
\psfrag{x01}[t][t]{0}%
\psfrag{x02}[t][t]{50}%
\psfrag{x03}[t][t]{100}%
\psfrag{x04}[t][t]{150}%
\psfrag{x05}[t][t]{200}%
%
% yticklabels:
\psfrag{v01}[r][r]{0}%
\psfrag{v02}[r][r]{2}%
\psfrag{v03}[r][r]{0}%
\psfrag{v04}[r][r]{2}%
\psfrag{v05}[r][r]{0}%
\psfrag{v06}[r][r]{2}%
\psfrag{v07}[r][r]{0}%
\psfrag{v08}[r][r]{2}%
\psfrag{v09}[r][r]{0}%
\psfrag{v10}[r][r]{2}%
\psfrag{v11}[r][r]{0}%
\psfrag{v12}[r][r]{2}%
\psfrag{v13}[r][r]{0}%
\psfrag{v14}[r][r]{2}%
\psfrag{v15}[r][r]{0}%
\psfrag{v16}[r][r]{2}%
\psfrag{v17}[r][r]{0}%
\psfrag{v18}[r][r]{2}%
%
% Figure:
%\resizebox{12cm}{!}{\includegraphics{laval_stability.eps}}%
\begin{figure}[!t]
    \begin{centering}
		\includegraphics[width=0.9\columnwidth]{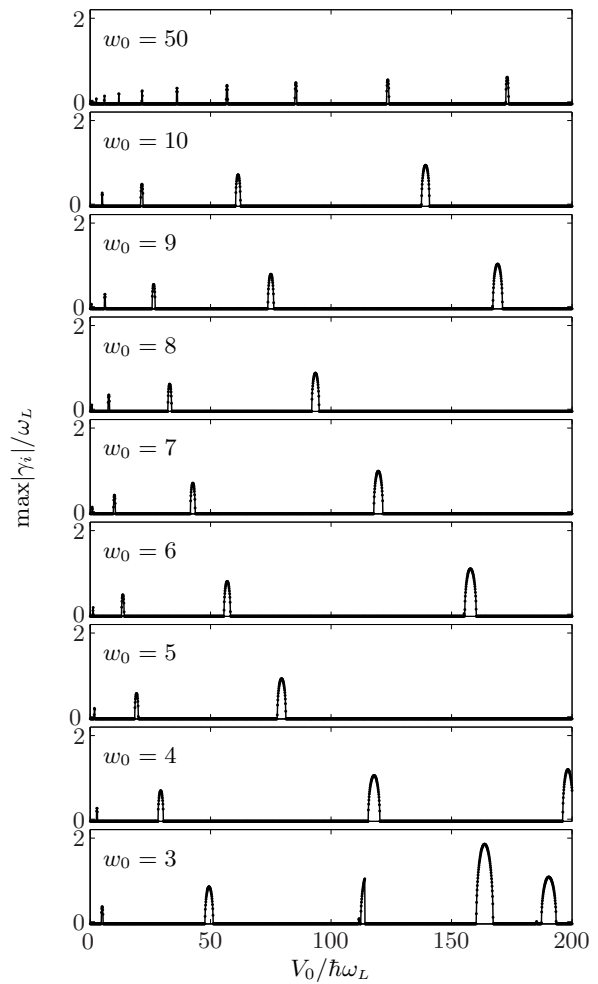}\par
		\end{centering}
		\vspace*{-10pt}
		\caption{\label{fig:stability}\textbf{Stability diagram:} We plot ${\rm max}\{|\gamma_i|\}$ against $V_0$ for stationary solutions with toroidal flow over a range of winding numbers. The quantum de Laval nozzle is unstable at the narrow spikes in the imaginary eigenvalues (which are equally narrow on a logarithmic scale); elsewhere the flow is stable.} 
		\vspace*{-10pt}
\end{figure}
\end{psfrags}%
%
% End laval_stability.tex

%===========================================
Figure \ref{fig:stability} shows the stability diagram for the QdLN that results by performing the diagonalization for a range of parameters, $V_0$ and $w_0$. For each point we have calculated the maximum of the absolute value for the imaginary part of all eigenvalues. The essential features we observe are: (i) there are regions exhibiting dynamic instabilities; (ii) these regions become narrower and smaller in magnitude, but more closely spaced for larger values of the winding number $w_0$, whereas they become broader and larger in magnitude as the potential depth $V_0$ increases. 

% This file is generated by the MATLAB m-file laprint.m. It can be included
% into LaTeX documents using the packages graphicx, color and psfrag.
% It is accompanied by a postscript file. A sample LaTeX file is:
%    \documentclass{article}\usepackage{graphicx,color,psfrag}
%    \begin{document}\input{modes_V01392_w10}\end{document}
% See http://www.mathworks.de/matlabcentral/fileexchange/loadFile.do?objectId=4638
% for recent versions of laprint.m.
%
% created by:           LaPrint version 3.16 (13.9.2004)
% created on:           08-Feb-2007 17:29:44
% eps bounding box:     15 cm x 20.5161 cm
% comment:              
%
\begin{psfrags}%
\psfragscanon%
%
% text strings:
\psfrag{s91}[b][b]{\color[rgb]{0,0,0}\setlength{\tabcolsep}{0pt}\begin{tabular}{c}(a)\end{tabular}}%
\psfrag{s92}[l][l]{\color[rgb]{0,0,0}\setlength{\tabcolsep}{0pt}\begin{tabular}{l}$i=1$\end{tabular}}%
\psfrag{s93}[b][b]{\color[rgb]{0,0,0}\setlength{\tabcolsep}{0pt}\begin{tabular}{c}(b)\end{tabular}}%
\psfrag{s94}[b][b]{\color[rgb]{0,0,0}\setlength{\tabcolsep}{0pt}\begin{tabular}{c}(c)\end{tabular}}%
\psfrag{s95}[l][l]{\color[rgb]{0,0,0}\setlength{\tabcolsep}{0pt}\begin{tabular}{l}$i=2$\end{tabular}}%
\psfrag{s96}[l][l]{\color[rgb]{0,0,0}\setlength{\tabcolsep}{0pt}\begin{tabular}{l}$i=3$\end{tabular}}%
\psfrag{s97}[l][l]{\color[rgb]{0,0,0}\setlength{\tabcolsep}{0pt}\begin{tabular}{l}$i=4$\end{tabular}}%
\psfrag{s98}[l][l]{\color[rgb]{0,0,0}\setlength{\tabcolsep}{0pt}\begin{tabular}{l}$i=5$\end{tabular}}%
\psfrag{s99}[l][l]{\color[rgb]{0,0,0}\setlength{\tabcolsep}{0pt}\begin{tabular}{l}$i=6$\end{tabular}}%
\psfrag{100}[l][l]{\color[rgb]{0,0,0}\setlength{\tabcolsep}{0pt}\begin{tabular}{l}$i=7$\end{tabular}}%
\psfrag{101}[t][t]{\color[rgb]{0,0,0}\setlength{\tabcolsep}{0pt}\begin{tabular}{c}$x/L$\end{tabular}}%
\psfrag{102}[l][l]{\color[rgb]{0,0,0}\setlength{\tabcolsep}{0pt}\begin{tabular}{l}$i=8$\end{tabular}}%
\psfrag{103}[t][t]{\color[rgb]{0,0,0}\setlength{\tabcolsep}{0pt}\begin{tabular}{c}$x/L$\end{tabular}}%
\psfrag{104}[t][t]{\color[rgb]{0,0,0}\setlength{\tabcolsep}{0pt}\begin{tabular}{c}$x/L$\end{tabular}}%
%
% xticklabels:
\psfrag{x01}[t][t]{-0.5}%
\psfrag{x02}[t][t]{0}%
\psfrag{x03}[t][t]{0.5}%
\psfrag{x04}[t][t]{-0.5}%
\psfrag{x05}[t][t]{0}%
\psfrag{x06}[t][t]{0.5}%
\psfrag{x07}[t][t]{-0.5}%
\psfrag{x08}[t][t]{0}%
\psfrag{x09}[t][t]{0.5}%
%
% yticklabels:
\psfrag{v01}[r][r]{-2}%
\psfrag{v02}[r][r]{0}%
\psfrag{v03}[r][r]{2}%
\psfrag{v04}[r][r]{-2}%
\psfrag{v05}[r][r]{0}%
\psfrag{v06}[r][r]{2}%
\psfrag{v07}[r][r]{0}%
\psfrag{v08}[r][r]{5}%
\psfrag{v09}[r][r]{10}%
\psfrag{v10}[r][r]{-2}%
\psfrag{v11}[r][r]{0}%
\psfrag{v12}[r][r]{2}%
\psfrag{v13}[r][r]{-2}%
\psfrag{v14}[r][r]{0}%
\psfrag{v15}[r][r]{2}%
\psfrag{v16}[r][r]{0}%
\psfrag{v17}[r][r]{5}%
\psfrag{v18}[r][r]{10}%
\psfrag{v19}[r][r]{-2}%
\psfrag{v20}[r][r]{0}%
\psfrag{v21}[r][r]{2}%
\psfrag{v22}[r][r]{-2}%
\psfrag{v23}[r][r]{0}%
\psfrag{v24}[r][r]{2}%
\psfrag{v25}[r][r]{0}%
\psfrag{v26}[r][r]{5}%
\psfrag{v27}[r][r]{10}%
\psfrag{v28}[r][r]{-2}%
\psfrag{v29}[r][r]{0}%
\psfrag{v30}[r][r]{2}%
\psfrag{v31}[r][r]{-2}%
\psfrag{v32}[r][r]{0}%
\psfrag{v33}[r][r]{2}%
\psfrag{v34}[r][r]{0}%
\psfrag{v35}[r][r]{5}%
\psfrag{v36}[r][r]{10}%
\psfrag{v37}[r][r]{-2}%
\psfrag{v38}[r][r]{0}%
\psfrag{v39}[r][r]{2}%
\psfrag{v40}[r][r]{-2}%
\psfrag{v41}[r][r]{0}%
\psfrag{v42}[r][r]{2}%
\psfrag{v43}[r][r]{0}%
\psfrag{v44}[r][r]{5}%
\psfrag{v45}[r][r]{10}%
\psfrag{v46}[r][r]{-2}%
\psfrag{v47}[r][r]{0}%
\psfrag{v48}[r][r]{2}%
\psfrag{v49}[r][r]{-2}%
\psfrag{v50}[r][r]{0}%
\psfrag{v51}[r][r]{2}%
\psfrag{v52}[r][r]{0}%
\psfrag{v53}[r][r]{5}%
\psfrag{v54}[r][r]{10}%
\psfrag{v55}[r][r]{-2}%
\psfrag{v56}[r][r]{0}%
\psfrag{v57}[r][r]{2}%
\psfrag{v58}[r][r]{-2}%
\psfrag{v59}[r][r]{0}%
\psfrag{v60}[r][r]{2}%
\psfrag{v61}[r][r]{0}%
\psfrag{v62}[r][r]{5}%
\psfrag{v63}[r][r]{10}%
\psfrag{v64}[r][r]{-2}%
\psfrag{v65}[r][r]{0}%
\psfrag{v66}[r][r]{2}%
\psfrag{v67}[r][r]{-2}%
\psfrag{v68}[r][r]{0}%
\psfrag{v69}[r][r]{2}%
\psfrag{v70}[r][r]{0}%
\psfrag{v71}[r][r]{5}%
\psfrag{v72}[r][r]{10}%
%
% Figure:
%\resizebox{12cm}{!}{\includegraphics{modes_V01392_w10.eps}}%
\begin{figure}[!t]
    \begin{center}
		\includegraphics[width=1\columnwidth]{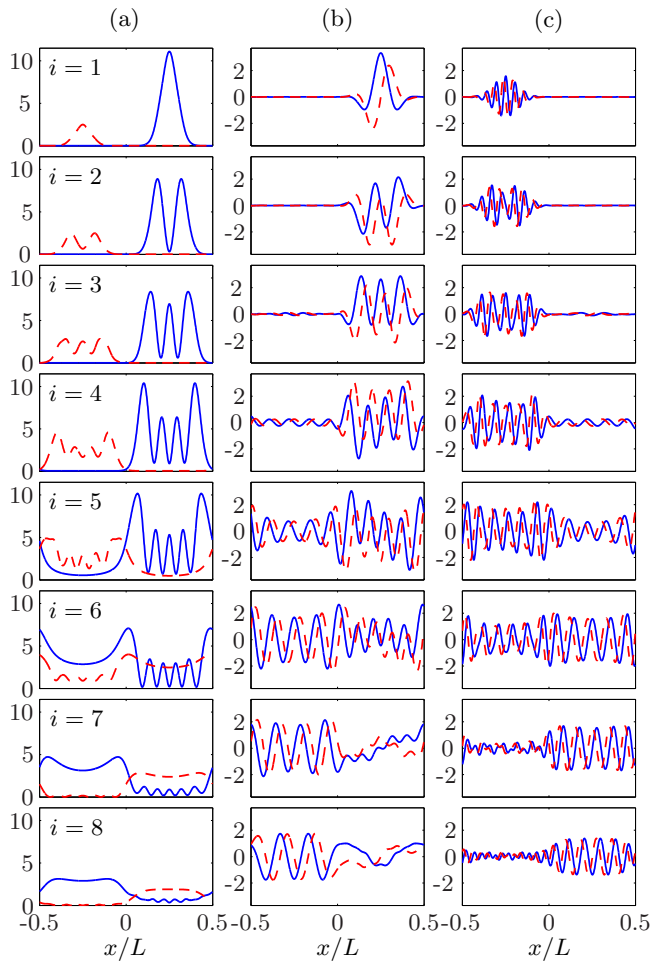}
		\caption{(Color online) Orthogonal mode functions for dynamically unstable configuration with $V_0 = 139.2.7\hbar\wL$ and $w_0 = 10$. Modes $1 \leq i \leq 8$ are shown. In column (a) $|U_i(x)|^2$ is given by the solid blue curve, and $|V_i(x)|^2$ is given by the dashed red curve. In the column (b) $\textrm{Re}(U_i(x))$ is given by the solid blue curve, and $\textrm{Im}(U_i(x))$ is given by the dashed red curve. In column (c) $\textrm{Re}(V^*_i(x))$ is given by the solid blue curve, and $\textrm{Im}(V^*_i(x))$ is given by the dashed red curve.} 
		\label{fig:modes_w10_V1392}
		\end{center}
\end{figure}
\end{psfrags}%
%
% End modes_V01392_w10.tex

% This file is generated by the MATLAB m-file laprint.m. It can be included
% into LaTeX documents using the packages graphicx, color and psfrag.
% It is accompanied by a postscript file. A sample LaTeX file is:
%    \documentclass{article}\usepackage{graphicx,color,psfrag}
%    \begin{document}\input{xMean}\end{document}
% See http://www.mathworks.de/matlabcentral/fileexchange/loadFile.do?objectId=4638
% for recent versions of laprint.m.
%
% created by:           LaPrint version 3.16 (13.9.2004)
% created on:           01-May-2007 14:47:54
% eps bounding box:     15 cm x 7.9503 cm
% comment:              
%
\begin{psfrags}%
\psfragscanon%
%
% text strings:
\psfrag{s01}[t][t]{\color[rgb]{0,0,0}\setlength{\tabcolsep}{0pt}\begin{tabular}{c}mode\end{tabular}}%
\psfrag{s02}[b][b]{\color[rgb]{0,0,0}\setlength{\tabcolsep}{0pt}\begin{tabular}{c}$\langle x  \rangle / L$\end{tabular}}%
%
% xticklabels:
\psfrag{x01}[t][t]{0}%
\psfrag{x02}[t][t]{0.1}%
\psfrag{x03}[t][t]{0.2}%
\psfrag{x04}[t][t]{0.3}%
\psfrag{x05}[t][t]{0.4}%
\psfrag{x06}[t][t]{0.5}%
\psfrag{x07}[t][t]{0.6}%
\psfrag{x08}[t][t]{0.7}%
\psfrag{x09}[t][t]{0.8}%
\psfrag{x10}[t][t]{0.9}%
\psfrag{x11}[t][t]{1}%
\psfrag{x12}[t][t]{0}%
\psfrag{x13}[t][t]{5}%
\psfrag{x14}[t][t]{10}%
\psfrag{x15}[t][t]{15}%
\psfrag{x16}[t][t]{20}%
%
% yticklabels:
\psfrag{v01}[r][r]{0}%
\psfrag{v02}[r][r]{0.1}%
\psfrag{v03}[r][r]{0.2}%
\psfrag{v04}[r][r]{0.3}%
\psfrag{v05}[r][r]{0.4}%
\psfrag{v06}[r][r]{0.5}%
\psfrag{v07}[r][r]{0.6}%
\psfrag{v08}[r][r]{0.7}%
\psfrag{v09}[r][r]{0.8}%
\psfrag{v10}[r][r]{0.9}%
\psfrag{v11}[r][r]{1}%
\psfrag{v12}[r][r]{-0.2}%
\psfrag{v13}[r][r]{-0.1}%
\psfrag{v14}[r][r]{0}%
\psfrag{v15}[r][r]{0.1}%
\psfrag{v16}[r][r]{0.2}%
\psfrag{v17}[r][r]{0.3}%
\psfrag{v18}[r][r]{0.4}%
\psfrag{v19}[r][r]{0.5}%
%
% Figure:
\begin{figure}[!t]
    \begin{centering}
		\includegraphics[width=0.9\columnwidth]{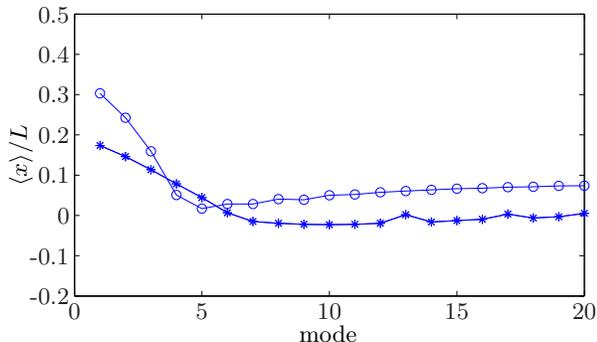}\par
		\end{centering}
		\vspace*{-10pt}
		\caption{\label{fig:xMean} Plot of $\langle x \rangle$ for the
		first 20 modes for two unstable cases, $w_0=3,
		V_0=163.7\hbar\wL$ (o) and
		$w_0=10, V_0=139.2\hbar\wL$ (*).} 
		\vspace*{-10pt}
\end{figure}
\end{psfrags}%
%
% End xMean.tex

In Fig.~\ref{fig:modes_w10_V1392} we show the corresponding mode functions $U_i(x)$ and $V_i(x)$ that result from the solutions of the BdG equations for the parameters $w_0 = 10$ and $V_0 = 139.2 \hbar \wL$, which has a dynamical instability for the modes $i = 5$, $6$. We note, although not shown here, the mode functions for $w_0 = 10$ and $V_0 = 100 \hbar \wL$ (dynamically stable) are very similar to the $V_0 = 139.2 \hbar \wL$ case.

The mode functions exhibit a rich structure for the low energy modes, not least being the sort of ``localization'' of modes that is associated with the acoustic black hole geometries. In particular, for modes $i \leq 4$, $U_i(x)$ is localized in the region $0 \leq x \leq 0.5$, which corresponds to the supersonic region, whereas $V_i(x)$ is localized in the region $-0.5 \leq x \leq 0$, corresponding to the subsonic region. For modes $i \geq 7$ we find the reverse is true in general, although the localization occurs to a lesser extent. Modes $i = 5$, $6$ (the dynamically unstable modes) indicate a crossover between these two regimes.
In Fig.~\ref{fig:xMean} we show the
average position of the quasiparticle modes for two unstable cases for comparison. The mean quasiparticle position for each mode is
calculated as~\cite{Isoshima2003}
$\langle x\rangle_n=(\langle x\rangle_{u_n}+\langle x\rangle_{v_n}-\langle x\rangle_{\phi_0})/\int dx\;
|u_n|^2+|u_n|^2$, where $\langle x\rangle_{u_n}=\int dx\;u_n^*(x)x u_n(x)$ with $\langle x\rangle_{v_n}$ defined
similarly, and $\langle x\rangle_{\phi_0}=\int dx\;\phi_0(x)^*x\phi_0(x)/\int dx\;|\phi_0(x)|^2$. We note that the negative energy modes (1-3 for $w_0=3$ and 1-5 for $w_0=10$)
are always located significantly inside the sonic horizon $\langle x \rangle >0$, while higher energy modes become located nearer the horizon.

\section{Dynamics}\label{sect:lavaldynamics}
To investigate the time dynamics of the QdLN we will use the stationary states as our starting point for quantum field theory simulations using the truncated Wigner method. 
We will compare the quasiparticle population dynamics in both stable and unstable regimes. First we briefly discuss the connection between our Bogoliubov analysis and the analogue Hawking effect.
\subsection{Two-mode Bogoliubov model}
The interaction Hamiltonian for a pair of dynamically unstable modes with $\omega_i=-\omega_j$, and with $|\gamma_i|=|\gamma_j|=\gamma$ is
\begin{equation}
H_{\rm int}=i\hbar\gamma[\hat{b}_- \hat{b}_+ - \hat{b}_-^\dag \hat{b}_+^\dag]
\end{equation}
which describes the formation of a two-mode squeezed state.
By finding the time evolution for the two mode density matrix according to $H_{\rm int}$ and averaging over the negative energy mode we obtain the density operator for the positive energy mode
\begin{equation}
\hat{\rho}_+(t)=\frac{1}{\cosh^2{\gamma t}}\sum_{n=0}^\infty(\tanh^2{\gamma t})^n\; |n\rangle_+\;_+\langle n|,
\end{equation}
a thermal state with mean occupation $\langle n\rangle_+ = \sinh^2{\gamma t}$. We thus have a loose analogy with the Hawking effect: pairs of quasiparticles can be produced with no
energy cost, such that one quasiparticle enters the negative energy state which is located inside the supersonic region (for our $w_0=10$ case in Fig. (\ref{fig:modes_w10_V1392}) this is mode 1), and the other is promoted to positive
energy (mode 6), which is centered much closer to the horizon. Tracing over the negative partner gives a thermal state for the postive energy mode. This connection has been
pointed out previously~\cite{Leonhardt2003}, but to our knowledge it has not been confirmed for
the trapped Bose gas using analysis of the GPE as we are able to do here (see Fig.~\ref{fig:twoModeComp}). However, we note that 
the analogy with the gravitational Hawking effect is incomplete, as may be seen from the time dynamics of the positive energy mode occupation number. A more direct Hawking analogue would generate a time 
independent multimode thermal emission spectrum, whereas here we have an exponentially growing, single mode emission. 

Thus far, we have found the elementary excitations for the QdLN and found that this indicates dynamically unstable configurations for certain sets of parameters. In order to
verify that such configurations do indeed lead to exponential growth in the unstable modes, we
consider the dynamics of the system. To do this we use the truncated Wigner method to perform short time
simulations of the full interacting quantum field theory describing the trapped Bose gas.

\subsection{Truncated Wigner method}
The Wigner representation provides a symmetrically ordered formalism for phase space simulations of quantum field theory. Symmetrically ordered operator averages are computed by
ensemble averaging many classical field trajectories. 
The truncated Wigner method~\cite{Steel1998,Sinatra2000,Sinatra2001,Sinatra2002,SGPEI,SGPEII,Polkovnikov2003} involves neglecting intractable third order derivatives in the equation of motion for the Wigner distribution. The method then reduces to numerically evolving
a multimode classical field using the GPE (\ref{gpelaval})~\cite{Steel1998}. The theory differs from pure mean field theory in that statistical fluctuations in the initial state
reproduce quantum fluctuations in the observables
extracted by ensemble averaging. The method is known to be accurate for short evolution times~\cite{Steel1998}. In the low temperature regime $k_B T\sim \epsilon_i$, the initial field is given by 
\begin{eqnarray}\label{psiwigner}
\psi(x, t = 0) = \psi_0(x) + \sum_{i > 0} \left[ U_{i}(x) \beta_{i} + 
V^*_{i}(x) \beta^*_{i} \right]
\end{eqnarray}
where $\psi_0(x)$ is a stationary state of the GPE (for our purposes, the transonic solutions of the QdLN), and where $U_{i}$ and $V_{i}$ are the Bogoliubov mode amplitudes of the
system. The complex random variables $\beta_i$ are
constructed as $\beta_i(t=0)=(\eta_1+i \eta_2)/\sqrt{2}$, where $\eta_i$ are real, normal Gaussian variates with $\overline{\eta_i}=0$ and
$\overline{\eta_i\eta_j}=\delta_{ij}(\bar{n}_i+1/2)$, and $\bar{n}_i=(e^{\epsilon_i/k_BT}-1)^{-1}$ is the thermal quasiparticle occupation. The notation $\overline{\eta}$ represents the
stochastic average over many samples of $\eta$. In this work we restrict our attention
to the zero temperature case to investigate the stability and dyanamics of our stationary solutions in the presence of vacuum fluctuations.

We expect the details of the quantum dynamics to depend sensitively on any instabilities, and indeed, 
according to the two mode model, instabilities can generate squeezing. We use the quasiparticle occupation numbers to look for confirmation of this effect in our simulations.
Dynamically, the Bogoliubov amplitudes can be extracted from the classical field as
\begin{eqnarray}
\beta_i(t) = \int d x \big( U_i^*(x) \psi(x, t) - V_i^*(x) \psi^*(x, t) \big)
\end{eqnarray}
which we use to monitor the populations during our simulations.
The quasiparticle number in each mode is then
\begin{eqnarray}
N_i(t) =\langle \hat{b}_i^\dag \hat{b}_i\rangle= \overline{\beta^*_i(t) \beta_i(t)} - \frac{1}{2}
\end{eqnarray}
where the bar indicates an ensemble average over many Wigner trajectories. 

In any simulations we must use a restricted basis, and our GPE evolution is numerically projected at each time step to ensure that the
system remains in the low energy subspace determined by our energy cut-off. Formally we are using the projected GPE (PGPE)~\cite{Davis2001b} to ensure consistent evolution of our restricted phase space.
For the evolution of the PGPE we have used the fourth-order Runge-Kutta in the interaction picture (RK4IP) algorithm \cite{BMCDThesis}, adapted to project into the low energy subspace defined by our momentum cutoff at $E_C=\hbar^2K_C^2/2m$~\cite{Davis2002}.
For the simulations presented here we have use $N_0 = 10^7$ condensate atoms, and $M = 1024$ modes for the system in the low energy subspace, corresponding to a dimensionless momentum cutoff
$K_C=2\pi M/L=6.4\times 10^3$. For all simulations we have used a time step for the RK4IP algorithm ensuring the change in total field normalization during each trajectory was $\Delta N/N < 10^{-6}$. 
\subsection{Results}

% This file is generated by the MATLAB m-file laprint.m. It can be included
% into LaTeX documents using the packages graphicx, color and psfrag.
% It is accompanied by a postscript file. A sample LaTeX file is:
%    \documentclass{article}\usepackage{graphicx,color,psfrag}
%    \begin{document}\input{bogmodes_w3_V163.7_unstable_tf1}\end{document}
% See http://www.mathworks.de/matlabcentral/fileexchange/loadFile.do?objectId=4638
% for recent versions of laprint.m.
%
% created by:           LaPrint version 3.16 (13.9.2004)
% created on:           24-Apr-2007 14:12:42
% eps bounding box:     15 cm x 11.25 cm
% comment:              
%
\begin{psfrags}%
\psfragscanon%
%
% text strings:
\psfrag{s02}[lt][lt]{\color[rgb]{0,0,0}\setlength{\tabcolsep}{0pt}\begin{tabular}{l}$\omega_L t$\end{tabular}}%
\psfrag{s03}[rt][rt]{\color[rgb]{0,0,0}\setlength{\tabcolsep}{0pt}\begin{tabular}{r}mode\end{tabular}}%
\psfrag{s04}[b][b]{\color[rgb]{0,0,0}\setlength{\tabcolsep}{0pt}\begin{tabular}{c}$N_i$\end{tabular}}%
%
% xticklabels:
\psfrag{x01}[t][t]{0}%
\psfrag{x02}[t][t]{0.5}%
\psfrag{x03}[t][t]{1}%
%
% yticklabels:
\psfrag{v01}[r][r]{1}%
\psfrag{v02}[r][r]{2}%
\psfrag{v03}[r][r]{3}%
\psfrag{v04}[r][r]{4}%
\psfrag{v05}[r][r]{5}%
\psfrag{v06}[r][r]{6}%
\psfrag{v07}[r][r]{7}%
\psfrag{v08}[r][r]{8}%
\psfrag{v09}[r][r]{9}%
\psfrag{v10}[r][r]{10}%
%
% zticklabels:
\psfrag{z01}[r][r]{0}%
\psfrag{z02}[r][r]{2}%
\psfrag{z03}[r][r]{4}%
\psfrag{z04}[r][r]{6}%
\psfrag{z05}[r][r]{8}%
\psfrag{z06}[r][r]{10}%
%
% Figure:
%\resizebox{12cm}{!}{\includegraphics{bogmodes_w3_V163.7_unstable_tf1.eps}}%
\begin{figure}[!ht]
    \begin{center}
		\includegraphics[width=0.9\columnwidth]{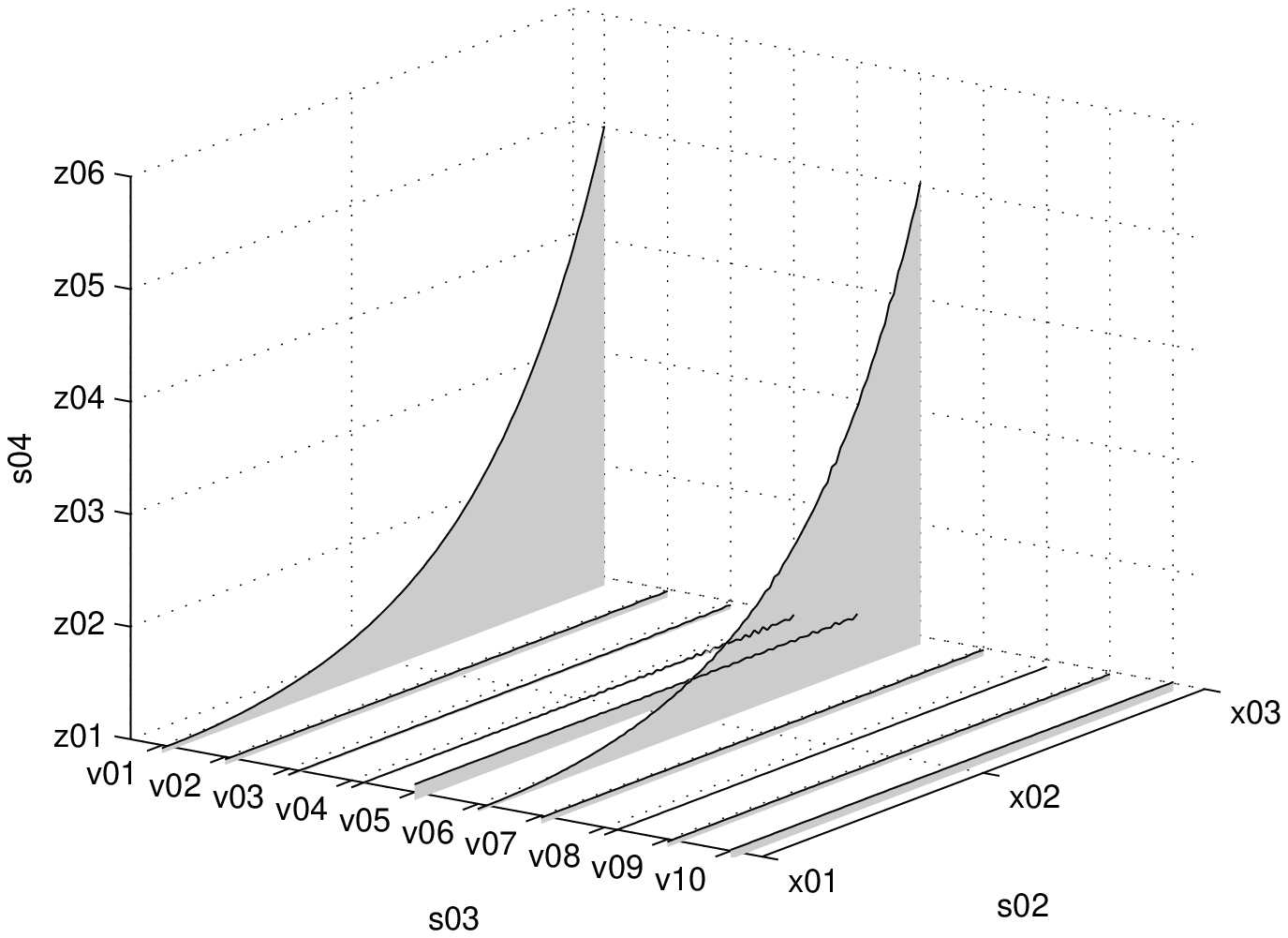}
		\caption{Bogoliubov mode populations for single trajectory from TWA evolution for parameters $V_0 = 163.7 \hbar \wL$, $w_0 = 3$, corresponding to a dynamically unstable configuration.} 
		\label{fig:lavaldynamics2}
		\end{center}
\end{figure}
\end{psfrags}%
%
% End bogmodes_w3_V163.7_unstable_tf1.tex

% generated by laprint.m
%
\begin{psfrags}%
\psfragscanon%
%
% text strings:
\psfrag{s01}[t][t]{\color[rgb]{0,0,0}\setlength{\tabcolsep}{0pt}\begin{tabular}{c}$\omega_{\scriptscriptstyle L} t$\end{tabular}}%
\psfrag{s02}[b][b]{\color[rgb]{0,0,0}\setlength{\tabcolsep}{0pt}\begin{tabular}{c}$N_6$\end{tabular}}%
%
% xticklabels:
\psfrag{x01}[t][t]{0}%
\psfrag{x02}[t][t]{0.1}%
\psfrag{x03}[t][t]{0.2}%
\psfrag{x04}[t][t]{0.3}%
\psfrag{x05}[t][t]{0.4}%
\psfrag{x06}[t][t]{0.5}%
\psfrag{x07}[t][t]{0.6}%
\psfrag{x08}[t][t]{0.7}%
\psfrag{x09}[t][t]{0.8}%
\psfrag{x10}[t][t]{0.9}%
\psfrag{x11}[t][t]{1}%
\psfrag{x12}[t][t]{0}%
\psfrag{x13}[t][t]{0.2}%
\psfrag{x14}[t][t]{0.4}%
\psfrag{x15}[t][t]{0.6}%
\psfrag{x16}[t][t]{0.8}%
\psfrag{x17}[t][t]{1}%
%
% yticklabels:
\psfrag{v01}[r][r]{0}%
\psfrag{v02}[r][r]{0.1}%
\psfrag{v03}[r][r]{0.2}%
\psfrag{v04}[r][r]{0.3}%
\psfrag{v05}[r][r]{0.4}%
\psfrag{v06}[r][r]{0.5}%
\psfrag{v07}[r][r]{0.6}%
\psfrag{v08}[r][r]{0.7}%
\psfrag{v09}[r][r]{0.8}%
\psfrag{v10}[r][r]{0.9}%
\psfrag{v11}[r][r]{1}%
\psfrag{v12}[r][r]{0}%
\psfrag{v13}[r][r]{2}%
\psfrag{v14}[r][r]{4}%
\psfrag{v15}[r][r]{6}%
\psfrag{v16}[r][r]{8}%
\psfrag{v17}[r][r]{10}%
%
% Figure:
\begin{figure}[!ht]
    \begin{centering}
		\includegraphics[width=\columnwidth]{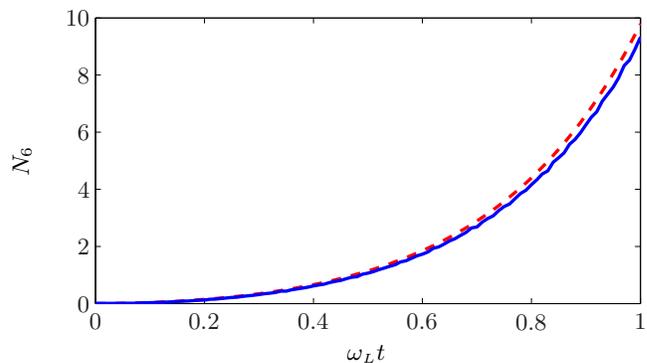}\par
		\end{centering}
		\vspace*{-10pt}
		\caption{\label{fig:twoModeComp} Population of the positive energy mode (mode 6 in Fig.~\ref{fig:lavaldynamics3}, solid line) with the result from Bogoliubov theory $\langle
		n\rangle_+=\sinh^2 \gamma_6 t$ (dashed line).} 
		\vspace*{-10pt}
\end{figure}
\end{psfrags}%
%
% End twoModeComp.tex

% This file is generated by the MATLAB m-file laprint.m. It can be included
% into LaTeX documents using the packages graphicx, color and psfrag.
% It is accompanied by a postscript file. A sample LaTeX file is:
%    \documentclass{article}\usepackage{graphicx,color,psfrag}
%    \begin{document}\input{bogmodes_w3_V163.7_unstable_tf5}\end{document}
% See http://www.mathworks.de/matlabcentral/fileexchange/loadFile.do?objectId=4638
% for recent versions of laprint.m.
%
% created by:           LaPrint version 3.16 (13.9.2004)
% created on:           24-Apr-2007 14:28:46
% eps bounding box:     15 cm x 11.25 cm
% comment:              
%
\begin{psfrags}%
\psfragscanon%
%
% text strings:
\psfrag{s02}[lt][lt]{\color[rgb]{0,0,0}\setlength{\tabcolsep}{0pt}\begin{tabular}{l}$\omega_L t$\end{tabular}}%
\psfrag{s03}[rt][rt]{\color[rgb]{0,0,0}\setlength{\tabcolsep}{0pt}\begin{tabular}{r}mode\end{tabular}}%
\psfrag{s04}[b][b]{\color[rgb]{0,0,0}\setlength{\tabcolsep}{0pt}\begin{tabular}{c}$N_i$\end{tabular}}%
%
% xticklabels:
\psfrag{x01}[t][t]{0}%
\psfrag{x02}[t][t]{1}%
\psfrag{x03}[t][t]{2}%
\psfrag{x04}[t][t]{3}%
\psfrag{x05}[t][t]{4}%
\psfrag{x06}[t][t]{5}%
%
% yticklabels:
\psfrag{v01}[r][r]{1}%
\psfrag{v02}[r][r]{2}%
\psfrag{v03}[r][r]{3}%
\psfrag{v04}[r][r]{4}%
\psfrag{v05}[r][r]{5}%
\psfrag{v06}[r][r]{6}%
\psfrag{v07}[r][r]{7}%
\psfrag{v08}[r][r]{8}%
\psfrag{v09}[r][r]{9}%
\psfrag{v10}[r][r]{10}%
%
% zticklabels:
\psfrag{z01}[r][r]{0}%
\psfrag{z02}[r][r]{1}%
\psfrag{z03}[r][r]{2}%
\psfrag{z04}[r][r]{3}%
\psfrag{z05}[r][r]{4}%
\psfrag{z06}[r][r]{5}%
\psfrag{z07}[r][r]{6}%
\psfrag{z08}[r][r]{7\setlength{\unitlength}{1ex}\begin{picture}(0,0)\put(0.5,1.5){$\times 10^{4}$}\end{picture}}%
%
% Figure:
%\resizebox{12cm}{!}{\includegraphics{bogmodes_w3_V163.7_unstable_tf5.eps}}%
\begin{figure}[!ht]
    \begin{center}
		\includegraphics[width=0.9\columnwidth]{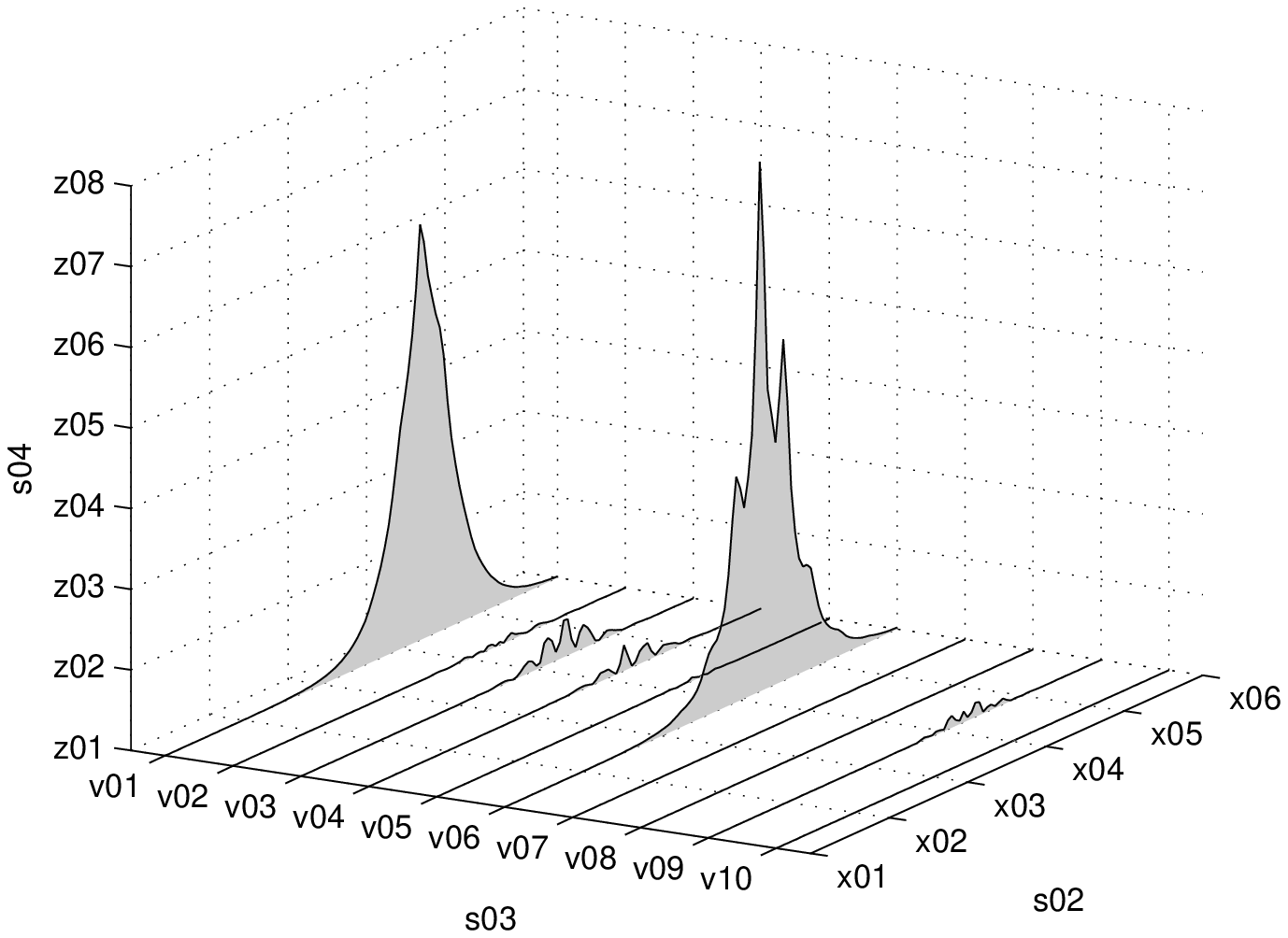}
		\caption{Bogoliubov mode populations for single trajectory from TWA evolution for parameters $V_0 = 163.7 \hbar \wL$, $w_0 = 3$, corresponding to a dynamically
		unstable configuration.} 
		\label{fig:lavaldynamics3}
		\end{center}
\end{figure}
\end{psfrags}%
%
% End bogmodes_w3_V163.7_unstable_tf5.tex

For a winding number of $w_0 = 3$, we have carried out time dynamical
simulations for 40 trajectories using the truncated Wigner method. The ensemble
averaged quasiparticle mode occupations have been calculated for two 
cases (refer to the stability diagram in Fig.~\ref{fig:stability}): (i) For the stable case $V_0
= 140 \hbar \wL$, the quasiparticle modes remain unnocupied during the interval $0\leq \wL t\leq
1$. This confirms the stability of the system to the extent possible given that our GPE stationary
state with vacuum noise is only an approximation to the true many body stationary state. (ii)
Fig.~\ref{fig:lavaldynamics2} shows the populations for the unstable case, $V_0 = 163.7 \hbar \wL$. In the latter case, modes $i = 1$, $6$ are unstable, 
and we observe exponential growth in these modes. The growth is seeded by the quantum vacuum fluctuations in the 
initial state and confirms the expectation that the system obeys the 
Hamiltonian (\ref{eq:bdgH2amplifier}) for a non-degenerate parametric amplifier at short times. 
Note that the mean quasiparticle occupation for modes $i > 10$ is also negligible compared with $i \leq 10$. In Fig.~\ref{fig:twoModeComp} we plot the population in the positive
energy mode of the unstable pair, $N_6$, which is seen to be in close agreement with the
dynamics expected from the Bogoliubov theory of Sec.~\ref{sect:lavalbog}.  

\begin{figure}[!t]
\centering
\subfigure[]{\raisebox{-0.5\height}[0.5\height][2.3cm]{%
\begin{psfrags}%
\psfragscanon%
%
% text strings:
\psfrag{s01}[b][b]{\color[rgb]{0,0,0}\setlength{\tabcolsep}{0pt}\begin{tabular}{c}$|\psi|^2/N_0$\end{tabular}}%
\psfrag{s02}[lt][lt]{\color[rgb]{0,0,0}\setlength{\tabcolsep}{0pt}\begin{tabular}{l}$\omega_L t$\end{tabular}}%
\psfrag{s03}[rt][rt]{\color[rgb]{0,0,0}\setlength{\tabcolsep}{0pt}\begin{tabular}{r}$x/L$\end{tabular}}%
%
% xticklabels:
\psfrag{x01}[t][t]{0}%
\psfrag{x02}[t][t]{1}%
\psfrag{x03}[t][t]{2}%
\psfrag{x04}[t][t]{3}%
\psfrag{x05}[t][t]{4}%
\psfrag{x06}[t][t]{5}%
%
% yticklabels:
\psfrag{v01}[r][r]{-0.5}%
\psfrag{v02}[r][r]{0}%
\psfrag{v03}[r][r]{0.5}%
%
% zticklabels:
\psfrag{z01}[r][r]{0}%
\psfrag{z02}[r][r]{1}%
\psfrag{z03}[r][r]{2}%
\includegraphics[width=0.45\columnwidth]{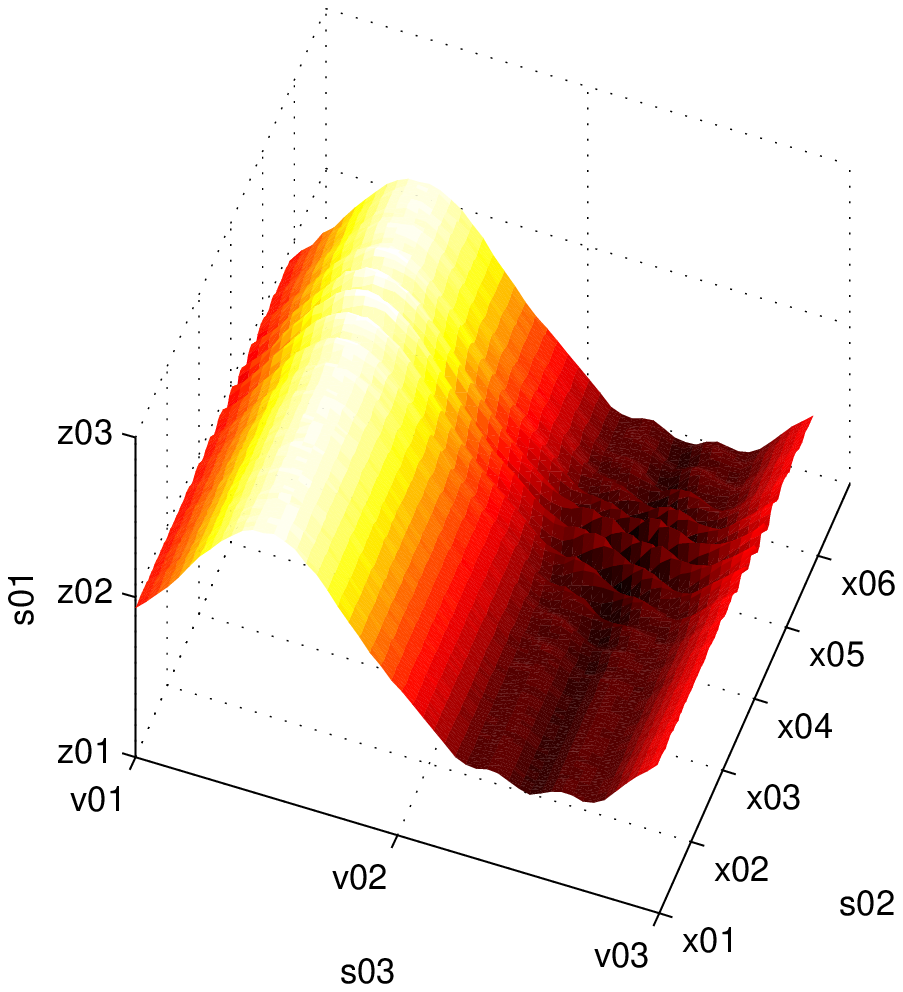}
\end{psfrags}%
}}
\hspace{3mm}    
\subfigure[]{\raisebox{-0.5\height}[0.5\height][2.3cm]{%
\begin{psfrags}%
\psfragscanon%
%
% text strings:
\psfrag{s05}[t][t]{\color[rgb]{0,0,0}\setlength{\tabcolsep}{0pt}\begin{tabular}{c} \\$\wL t$\end{tabular}}%
\psfrag{s06}[b][b]{\color[rgb]{0,0,0}\setlength{\tabcolsep}{0pt}\begin{tabular}{c}$x/L$\end{tabular}}%
\psfrag{s09}[][]{\color[rgb]{0,0,0}\setlength{\tabcolsep}{0pt}\begin{tabular}{c} \end{tabular}}%
\psfrag{s10}[][]{\color[rgb]{0,0,0}\setlength{\tabcolsep}{0pt}\begin{tabular}{c} \end{tabular}}%
\psfrag{s11}[b][b]{\color[rgb]{0,0,0}\setlength{\tabcolsep}{0pt}\begin{tabular}{c}$\Delta n$\end{tabular}}%
%
% xticklabels:
\psfrag{x01}[t][t]{0}%
\psfrag{x02}[t][t]{0.1}%
\psfrag{x03}[t][t]{0.2}%
\psfrag{x04}[t][t]{0.3}%
\psfrag{x05}[t][t]{0.4}%
\psfrag{x06}[t][t]{0.5}%
\psfrag{x07}[t][t]{0.6}%
\psfrag{x08}[t][t]{0.7}%
\psfrag{x09}[t][t]{0.8}%
\psfrag{x10}[t][t]{0.9}%
\psfrag{x11}[t][t]{1}%
\psfrag{x12}[t][t]{0}%
\psfrag{x13}[t][t]{0.1}%
\psfrag{x14}[t][t]{0.2}%
\psfrag{x15}[t][t]{0.3}%
\psfrag{x16}[t][t]{0.4}%
\psfrag{x17}[t][t]{0.5}%
\psfrag{x18}[t][t]{0.6}%
\psfrag{x19}[t][t]{0.7}%
\psfrag{x20}[t][t]{0.8}%
\psfrag{x21}[t][t]{0.9}%
\psfrag{x22}[t][t]{1}%
\psfrag{x23}[t][t]{0}%
\psfrag{x24}[t][t]{1}%
\psfrag{x25}[t][t]{2}%
\psfrag{x26}[t][t]{3}%
\psfrag{x27}[t][t]{4}%
\psfrag{x28}[t][t]{5}%
%
% yticklabels:
\psfrag{v01}[r][r]{0}%
\psfrag{v02}[r][r]{0.1}%
\psfrag{v03}[r][r]{0.2}%
\psfrag{v04}[r][r]{0.3}%
\psfrag{v05}[r][r]{0.4}%
\psfrag{v06}[r][r]{0.5}%
\psfrag{v07}[r][r]{0.6}%
\psfrag{v08}[r][r]{0.7}%
\psfrag{v09}[r][r]{0.8}%
\psfrag{v10}[r][r]{0.9}%
\psfrag{v11}[r][r]{1}%
\psfrag{v12}[l][l]{-0.1}%
\psfrag{v13}[l][l]{0}%
\psfrag{v14}[l][l]{0.1}%
\psfrag{v15}[l][l]{0.2}%
\psfrag{v16}[r][r]{-0.5}%
\psfrag{v17}[r][r]{0}%
\psfrag{v18}[r][r]{0.5}%
\includegraphics[width=0.45\columnwidth]{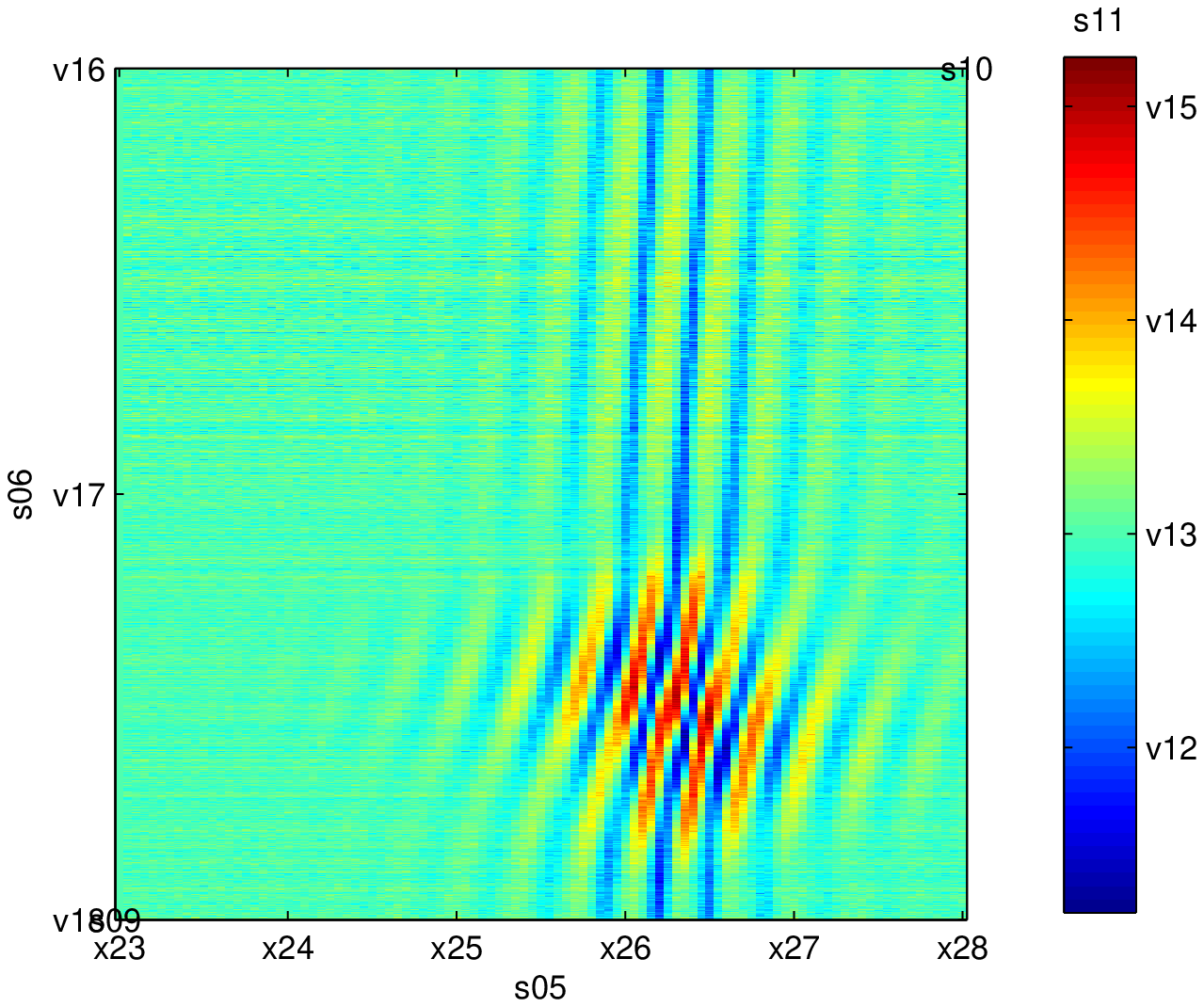}
\end{psfrags}%
}}
		\caption{(Color online) Coordinate space density vs time for unstable configuration with $w_0 = 3$ and $V_0 = 163.7 \hbar \wL$: (a) gives the normalized density of
		the field by $n(x, t) = |\psi(x, t)|^2/N_0$; (b) gives an intensity plot of the change in density from the initial state by $\Delta n(x, t) = n(x, t) - n(x, 0)$.
		In plot (a) the high frequency noise has been filtered out for clarity.}
		\label{fig:lavaldynamics4}
\end{figure}

We have also investigated the behaviour of the dynamically unstable configuration for longer times. Single trajectory results for a simulation time 
of $\omega_L t = 5$ are shown in Fig.~\ref{fig:lavaldynamics3}. Here we observe growth in the unstable modes ($i = 1$, $6$) until $\omega_L t \sim 3.25$ where there 
is a peak in the mode populations, followed by a decay of occupation numbers. 
Therefore the system undergoes a period of excitation followed by an apparent return to the initial unexcited state. This is also evident in the 
coordinate space density plots for the same simulation given in Figs.~\ref{fig:lavaldynamics4} (a) and (b). In particular, plot (a) shows large 
scale density fluctuations for $2 \lesssim t \lesssim 3.5$. Plot (b) shows the density relative to the initial state, from which it is clear that 
the density fluctuations are localised in the region $0 \leq x \leq 0.5$ corresponding to the supersonic region for the system.  It seems plausible 
that such excitations and revivals should continue to repeat, which would result in a ``ringing'' type excitation of the condensate. However, due to 
the significant computational time required, we did not check this prediction.  

The recurrence of the system is evidently due to nonlinear mode mixing, which is neglected in the BdG analysis. 
In particular, the back-reaction of quasiparticle modes on the condensate should become significant for large mode occupations. Moreover, the 
topological constraint imposed by the periodicity of the system (ie.~by a fixed winding number) means that the decay of circulation for the superfluid 
flow is forbidden. Evidently the decay channel for this instability is inhibited. The system cannot reach a quasi-stationary 
state corresponding to a different value of $w_0$, without a corresponding change in $V_0$ and the damping of topological charge; one mechanism for this process would be soliton
shedding, but we have not observed this in our simulations of this system. 

We also investigated the dynamics for the higher winding number $w_0 = 10$. 
In particular, we have examined two flows, which are (i) the dynamically stable (according to the BdG analysis) flow, $V_0 = 100 \hbar \wL$; 
and (ii) the dynamically unstable flow $V_0 = 139.2 \hbar \wL$. We briefly summarize the results for this case as we have found it to be typical of high winding number behaviour.
From our ensemble averaged time dynamics carried out in a similar manner to the previous cases, we found (1) that linear stability, evidenced by time-independent vacuum
quasiparticle population, was confirmed for short time quantum dynamics; (2) for the
unstable configuration, chaotic multimode dynamics is evident in the otherwise stable time interval, in particular, we did not observe simple two mode squeezing
dynamics of the kind seen for $w_0=3$. This behaviour is not 
unexpected since for the higher winding number, the effective nonlinearity required is very large (recall from the perturbation 
theory of Sec.~\ref{sect:lavalstationary} the nonlinearity scales with the square of winding number). In our quantum dynamical simulations elementary excitations interact
with each other, giving rise to Landau-Beliaev damping~\cite{Pitaevskii1997,Giorgini1998}, and this effect is more pronounced at higher nonlinearities.

%\section{Connection with Hawking effect}\label{sect:lavalhawking}
%\input{laval_hawkingconnection}

\section{Conclusions}\label{sect:lavalconclusions}
We have introduced and analyzed the quantum de Laval nozzle, a toroidal geometry for a BEC that exhibits both a black and white sonic horizon. 
Using hydrodynamic theory we have found transonic solutions, which we used to find transonic stationary solutions of the Gross-Pitaevskii equation. The qualitative properties of the GPE solutions are well described by hydrodynamic perturbation theory at lowest order. The system has broad dynamical instabilities for certain values of the winding number $w_0$ and potential depth $V_0$. 

We constructed normalizable Bogoliubov modes for the dynamical instabilities, which couple modes of positive and negative energy. This analysis leads to a two mode squeezing Hamiltonian
term corresponding to non-degenerate parametric amplification, which leads to exponential population growth of unstable mode occupation with time -- this represents the closest analogy with
the Hawking effect for our trapped quantum system.

To analyze this picture further, we have investigated the dynamics of several configurations using the truncated Wigner method which from an analogue model point of view includes
the effects of nonlinear interactions between modes and back reaction. For low winding number we observe non-degenerate parametric amplification type dynamics at the instability,
confirming the two-mode Bogoliubov model validity, while for the stable configuration there is negligible growth in all modes. 
%This confirms that higher order interactions which are neglected in the Bogoliubov quasiparticle picture become important for large winding number. 

From the stability analysis, we note for large winding number solutions: (i) the number of unstable regions increases, and they become narrower; (ii) the nonlinearity increases so that the system approaches the hydrodynamic regime; and (iii) short wavelength negative energy modes, for which the geometric acoustics approximation may be valid, increase in number. The combination of these effects indicates that in the limit of high winding number it may be possible to recover a classical fluid description of this system, for which the prediction of a thermal spectrum from Unruh \cite{Unruh1981} and Visser \cite{Visser1993} should be experimentally verifiable. 

In contrast, for the relatively low winding numbers we have considered here, quantum effects are significant and the semiclassical approximation breaks down. 
It apparently becomes necessary to revise the concept of the analogue Hawking effect for trapped Bose-Einstein condensates in this regime. 
%Nevertheless, the central feature of the Hawking process is still evident in our simulations: correlated pairs of positive and negative energy excitations are generated by an instability associated with the acoustic horizon.

%%%%%%%%%%%%%%%%%%%%%%%%%%%%%%%%%%%%%%%%%%%%%%%%%%%%%%%%%%%%%%%%%%%%%%%%%%%%%%%%%%%%%
\section*{Acknowledgements}
The authors would like to thank M.~Visser, S.~Weinfurtner, M.~K.~Olsen and C.~M.~Savage for useful discussions.
This research was supported by the Marsden Fund, the Tertiary Education Commission, Victoria University of Wellington, and the Australian Research Council.
%====================================================================================
%******************************************************************************
\bibliographystyle{prsty}
\bibliography{./refs}

\end{document}